\documentclass[12pt]{article}

\usepackage[totalwidth=460truept,totalheight=622truept]{geometry}
\usepackage{latexsym,graphicx,amsfonts,amssymb,amsmath,xcolor}
\usepackage[hypertexnames=false,hidelinks]{hyperref}

\global\arraycolsep=1truept

\begin{document}

\ \null

\vskip.0truecm

\begin{center}
{\huge \textbf{Purely Virtual Extension}}

\vskip.6truecm

{\huge \textbf{of Quantum Field Theory}}

\vskip.6truecm

{\huge \textbf{for\ Gauge Invariant Fields:}}

\vskip.6truecm

{\huge \textbf{Quantum Gravity}}

\vskip1truecm

\textsl{Damiano Anselmi}

\vskip.1truecm

{\small \textit{Dipartimento di Fisica \textquotedblleft
E.Fermi\textquotedblright , Universit\`{a} di Pisa, Largo B.Pontecorvo 3,
56127 Pisa, Italy}}

{\small \textit{INFN, Sezione di Pisa, Largo B. Pontecorvo 3, 56127 Pisa,
Italy}}

{\small damiano.anselmi@unipi.it}

\vskip1truecm

\textbf{Abstract}
\end{center}

Quantum gravity is extended to include purely virtual \textquotedblleft
cloud sectors\textquotedblright , which allow us to define a complete set of
point-dependent observables, including a gauge invariant metric and gauge
invariant matter fields, and calculate their off-shell correlation functions
perturbatively. The ordinary on-shell correlation functions and the $S$
matrix elements are unaffected. Each extra sector is made of a cloud field,
its anticommuting partner, a \textquotedblleft
cloud-fixing\textquotedblright\ function and a cloud Faddeev-Popov
determinant. The additional fields are purely virtual, to ensure that no
ghosts propagate. The extension is unitary. In particular, the off-shell,
diagrammatic version of the optical theorem holds. The one-loop two-point
functions of dressed scalars, vectors and gravitons are calculated. Their
absorptive parts are positive, cloud independent and gauge independent,
while they are unphysical if non purely virtual clouds are used. We
illustrate the differences between our approach to the problem of finding a
complete set of observables in quantum gravity and other approaches
available in the literature.

\vfill\eject

\section{Introduction}

\label{intro}\setcounter{equation}{0}

Defining point-dependent observables in general relativity is tricky,
because the coordinates are not physical quantities, but just
parametrizations of the location. A simple way out is available when the
spacetime point is associated with a matter distribution. Consider, for
definiteness, four scalar fields $\phi ^{i}(x)$, $i=1,\ldots 4$, and assume
that they depend on the coordinates $x^{\mu }$ in such a way that it is
possible to invert $x^{\mu }$ as functions $x^{\mu }(\phi )$ of $\phi ^{i}$.
Then, every further field, say a fifth scalar $\varphi (x)$, can be written
as a function of the reference fields $\phi ^{i}$: $\varphi (x)\rightarrow
\varphi (x(\phi ))\equiv \tilde{\varphi}(\phi )$. The function $\tilde{%
\varphi}(\phi )$ is obviously invariant under general changes of
coordinates. The basic idea is to go back to the physical location every
time we change coordinates. The $\phi ^{i}$ do not need to be new,
independent fields, but can be functions of the metric itself.

This line of thinking has been pursued in the literature for a long time 
\cite{precedent,Komar,clockclouds,Rovelli,Giddings}. In refs. \cite{Komar}
Komar and Bergmann use functions of the metric. In refs. \cite%
{clockclouds,Rovelli} the fields $\phi ^{i}$ describe physical matter.
Donnelly and Giddings \cite{Giddings} view them as functions of the metric,
for purposes similar to the Coulomb-Dirac dressing of QED \cite{Dirac}, the
Lavelle-McMullan dressing of non-Abelian gauge theories \cite{Lavelle}, the
worldline dressing and the Wilson lines.

Ultimately, the presence of the observer, which is \textquotedblleft
matter\textquotedblright , is what breaks general covariance, so we may want
to view the reference fields as independent matter. In this spirit, we have
to take into account that the fundamental theory is changed by the presence
of the fields $\phi ^{i}$. In ref. \cite{Rovelli} Rovelli questions the need
to change the physical world for this purpose and proposes an improvement
inspired by the GPS technology, based on a minimal amount of additional
matter. Yet, one still needs to provide the physics of the additional matter
and add it to the physics of the fundamental theory. Viewing the reference
fields as functions of the metric is more appealing from the conceptual
point of view, since it does not force us to leave the realm of pure gravity.

These problems are more challenging at the fundamental level, especially in
quantum gravity, where they concern our understanding of the fundamental
physics of nature.

In this paper we pursue a new strategy, which may shed a different light on
the issue. We introduce purely virtual \cite{PVP20,diagrammarMio}
independent fields $\zeta ^{\mu }$, which we call cloud fields, and their
anticommuting partners $H^{\mu }$. We extend quantum field theory to include
such fields perturbatively in quantum gravity (and general relativity)
without affecting the fundamental laws of physics.

The cloud fields play the roles of $\phi ^{i}$. Precisely, the fields $\phi
^{i}$ should be imagined as the differences $x^{\mu }-\zeta ^{\mu }(x)$. An
arrangement of this type also appears in \cite{Giddings}. In our approach,
however, the $\zeta ^{\mu }$ are neither additional matter, nor functions of
the metric, but independent fields, with their own (higher-derivative)
propagators, and their own interactions. Moreover, they must be accompanied
by anticommuting partners $H^{\mu }$ in a suitable way,\ and rendered purely
virtual, because only in that case the fundamental physics does not change,
and no extra degrees of freedom (which would include ghosts) are propagated.

The cloud fields $\zeta ^{\mu }$ are used to surround the elementary fields
of the theory, such as the metric tensor, with appropriate dressings, in
order to render them invariant under infinitesimal changes of coordinates.
The correlation functions of the dressed fields are new physical quantities,
and provide predictions that can in principle be tested experimentally.

The anticommuting partners $H^{\mu }$ are used to endow the extended theory
with a certain cloud symmetry, to ensure that the fundamental interactions
are unaffected by the presence of the cloud sectors. Specifically: $i$) the
correlation functions of the undressed fields are unchanged, and $ii$) the $%
S $ matrix amplitudes of the dressed fields coincide with the usual $S$
matrix amplitudes of the undressed fields. Once these goals are achieved, we
can concentrate on the new correlation functions. Moreover, we can view the
usual (undressed) fields as mere integration and diagrammatic tools, and use
dressed fields everywhere else. This way, gauge invariance and gauge
independence become manifest in every operation we make.

Because they are independent fields, $\zeta ^{\mu }$ and $H^{\mu }$ might be
viewed as a sort of matter. Then, however, their purely virtual nature makes
them \textquotedblleft fake matter\textquotedblright . Because they are
introduced to be ultimately projected away (that is to say, integrated out),
they might be understood as \textquotedblleft functions of everything
else\textquotedblright , at least in some particular cases (like the
classical limit). Nevertheless, they cannot be viewed as functions of the
other fields beyond the tree level, since they keep circulating in loops. In
view of these remarks, it is better to understand $\zeta ^{\mu }$ and $%
H^{\mu }$ as new entities, defined by the very same formalism we develop in
the paper.

Note that the dressed metric just propagates the two graviton helicities. In
the approaches of refs. \cite{clockclouds,Rovelli} it may propagate six
degrees of freedom (the additional ones coming from the four reference
scalars).

We achieve the goals we have stated in a fully perturbative regime. By
construction, the extended theory is local and unitary. Moreover, it is
renormalizable (if the underlying gravity theory is renormalizable), up to
the cloud sectors, which may be nonrenormalizable due to their
arbitrariness. Although the observables that we define are invariant under
infinitesimal changes of coordinates, they are not necessarily invariant
under global changes of coordinates. From the conceptual and physical points
of view, this is what we need: we break the global symmetry (our
observations do that most of the times) without violating unitarity.

In a parallel paper \cite{AbsoPhys}, we explore similar issues in gauge
theories.

Throughout the paper, \textquotedblleft on shell\textquotedblright\ means on
the mass shell, and refers to the $S$ matrix asymptotic states. The
correlation functions of the dressed fields differ from the usual
correlation functions anytime the external legs are not asymptotic states on
the mass shell. The word \textquotedblleft virtuality\textquotedblright\ is
used in connection with the concept of pure virtuality, and refers to the
removal of all the on-mass-shell contributions to the correlation functions
due to a given particle, preserving unitarity \cite{diagrammarMio}. That
particle is then called purely virtual.

The notion of pure virtuality relies on a new diagrammatics \cite%
{diagrammarMio}, which allows us to introduce particles that mediate
interactions without ever being on shell. The construction is compatible
with unitarity, and takes advantage of the possibility of splitting the
usual optical theorem \cite{unitarity} into independent, spectral optical
identities, associated with different (multi)thresholds \cite{diagrammarMio}%
. When we want to render certain particles $\chi $ purely virtual, and
calculate diagrams involving them, we need to start from ordinary Feynman
diagrams, as if $\chi $ were physical particles or ghosts, and remove the
contributions of the $\chi $-dependent nontrivial thresholds,\ as explained
in ref. \cite{diagrammarMio}. Since the spectral optical identities
involving those thresholds drop out altogether from the optical theorem,
unitarity is manifestly preserved, or enforced (if all the potential ghosts
are rendered purely virtual).

The removed degrees of freedom can also be understood as fake particles, or
\textquotedblleft fakeons\textquotedblright . The main application of this
concept is the formulation of a consistent theory of quantum gravity \cite%
{LWgrav}, which leads to observationally testable predictions in
inflationary cosmology \cite{ABP}. At the phenomenological level, fakeons
evade common constraints that preclude the usage of normal particles \cite%
{Tallinn1,Tallinn2}.

\medskip

Before plunging into the technical details, we illustrate some applications
of the results of this paper. Although the usual $S$ matrix amplitudes do
not change, we can define new types of scattering processes. Consider, for
example, \textquotedblleft short-distance\textquotedblright\ scattering
processes among gravitons (or quarks and gluons), that is to say, processes
that occur within distances such that the incoming and outgoing states are
not allowed to become free (interactions in a quark guon plasma,
interactions between quarks and gluons at distances that are smaller or much
smaller than the proton radius, interactions within the Plank scale, or some
slightly larger scales, in strongly interacting quantum gravity, etc.). We
cannot advocate the notion of asymptotic state to study such processes. In
our language, they do not need to be on the mass shell and, therefore, they
may be gauge dependent. The results of this paper show that we can actually
define these scattering processes by means of dressed fields. The price is
that, although the results we obtain are physical (i.e., they are gauge
invariant and obey the optical theorem), they depend on the clouds, through
the cloud parameters, which we denote by $\tilde{\lambda}$. These parameters
do not belong to the fundamental theory. Rather, they describe features of
the experimental setup (experimental resolutions, finite volume effects,
finite temperature effects, dependence on a background, etc.). The $\tilde{%
\lambda}$ dependence of the results means that it is impossible to eliminate
the influence of the observer on the observed phenomenon. Yet, we can get
rid of the $\tilde{\lambda}$ dependences at a second stage, by sacrificing a
few measurements to calibrate the instrumentation. Once the values of the
parameters $\tilde{\lambda}$ are determined, everything else is predicted
uniquely, and can be confirmed or falsified. Something similar occurs in the
study of infrared divergences of the $S$ matrix amplitudes in gauge and
gravity theories \cite{infred}, where it is necessary to specify the
resolution of the apparatus. The resolution is also necessary to describe
the observation of unstable particles, like the muon \cite{muon}, which do
not admit asymptotic states in a strict sense.

The dependence on the clouds is less surprising if we think that even the
gauge invariant correlation functions built by means of Wilson lines depend
on the Wilson lines themselves. Different Wilson lines describe situations
of different interest, experimentally. Yet, those dependences must be there,
since they are a reflection on the choices of the observer.

Another application concerns precisely the infrared behaviors of gauge and
gravity theories. The standard way to deal with this problem is to resum
another type of dressing, made of soft and collinear photons, gluons and
gravitons \cite{infred}. An alternative way to regularize the infrared
divergences is by going a little bit off the mass shell. Working with the
usual correlation functions, however, this operation violates gauge
invariance. The new correlation functions considered here, made of dressed
fields, allow us to go off shell without breaking the local symmetries, and
can provide an alternative way to probe the infrared behaviors of gauge and
gravity theories.

Before plunging into the topics of the paper, we point out some crucial
differences between our methods and purposes, and the ones of other
approaches, which may appear to have something in common with ours at first,
but are actually very different. We are referring to the Stueckelberg
formalism \cite{stueckelberg} and the compensator-field approach \cite%
{compensator}. The former is used to describe massive vectors, which is not
our goal here. The latter is used to rephrase the theory in a way that is
convenient for several applications, but is not meant to change the physical
cohomology.

In particular, after extending the theory, we still define the physical
observables as being gauge invariant: they are not required to be invariant
under the extra (cloud) transformations. This changes the cohomology of
physical observables, and makes the extension nontrivial. A possible source
of trouble, however, comes from the fact that the extension is
higher-derivative, and may inject unphysical degrees of freedom in the
theory, in the form of ghosts. What saves the day is the last ingredient of
our construction, that is to say, the requirement that the whole extension
be purely virtual. This makes the construction unitary, and essentially
different from the other proposals available in the literature. Moreover, it
guarantees that the fundamental spectrum of the theory is unmodified.

On these premises, we manage to build the gauge-invariant, cloud-dependent
fields. While the usual correlation functions do not change, we are able to
treat new physical correlation functions: those that contain insertions of
the gauge-invariant fields, such as point-dependent observables, including a
gauge invariant metric tensor, as well as gauge invariant matter fields. We
can also consider new scattering processes, like the short distance
processes mentioned above. We stress that the ones we study are not just
gauge invariant correlation functions: they are (non-asymptotic) correlation
functions of gauge invariant fields.

The explicit calculations we make show that if the extra fields are not
purely virtual, but quantized by means of the usual Feynman prescription,
for example, they cause disasters, by propagating ghosts. This is what marks
a crucial distinction between the construction of this paper and its
alternatives. In principle, one can form correlation functions of gauge
invariant fields in other approaches, including the compensator-field
method. Then, however, one must address the problem of ghosts, otherwise the
results turn out to be unphysical. This problem also plagues the correlation
functions of Wilson lines, as shown in ref. \cite{AbsoPhys}. It can only be
cured, as far as we know today, by resorting to a purely virtual extension.
Typically, issues with Lorentz invariance arise in approaches employing
Dirac clouds \cite{Dirac}, which are of the Coulomb type. These and other
problems are overcome, or addressed in much easier ways, in the approach of
this paper.

Everything works as long as we keep the usual sector and the cloud sectors
to some extent separated. We show it is indeed possible, because
renormalization preserves the unmixing. In particular, the functions that
define the clouds should be gauge invariant, while the usual gauge-fixing
functions should be cloud invariant. In this respect, we recall that
restrictions on the gauge-fixing choice are not unusual. A familiar one is
adopted in the context of the background field method, where the
gauge-fixing must be invariant under the background transformations,
otherwise it spoils the virtues of the method. Similarly, the choices we
make in our approach highlight a number of virtues that are more difficult
to uncover otherwise.

The approach of this paper can offer a better understanding of the physics
that lies beyond the realm of scattering processes in quantum field theory,
and provides an anwser to the\ problem of finding a complete set of
observables in quantum gravity.

Throughout the paper we work with the dimensional regularization \cite%
{dimreg}, $\varepsilon =4-D$ denoting the difference between the physical
dimension and the continued one.

The paper is organized as follows. In section \ref{cloudf} we introduce the
fields we need to build the cloud sectors. In section \ref{Bata} we recall
the standard Batalin-Vilkovisky formalism for gravity and the Zinn-Justin
master equation. In section \ref{BataCloud} we define the cloud sector. In
section \ref{cloudindep} we show that the ordinary correlation functions are
unaffected by the cloud sector, and so are the $S$ matrix amplitudes. In
section \ref{cloudcorre} we build the correlation functions of the dressed
fields. In section \ref{duality} we prove that the gauge-trivial sector of
the theory and the cloud sector are mirrored into one another by a certain
duality relation. In section \ref{multiclouds} we add several copies of the
could sector and show that each insertion, in a correlation function, can be
dressed with its own, independent cloud. In sections \ref{dressed2ptf} and %
\ref{pvc} we compute the one-loop two-point functions of the dressed
scalars, vectors and gravitons. We do so in Einstein gravity and in quantum
gravity with purely virtual particles. In section \ref{dressed2ptf} we work
with covariant clouds, and show that the absorptive parts are unphysical. In
section \ref{pvc} we turn to purely virtual clouds, and show that the
absorptive parts are then physical. In section \ref{renormalization} we
prove that the extended theory is renormalizable. Section contains \ref%
{conclusions} the conclusions.

\section{The cloud field, its anticommuting partner, and the dressed fields}

\setcounter{equation}{0}\label{cloudf}

In this section we lay out the basic notions that are needed to build the
cloud sectors. Let 
\begin{equation}
\delta _{\xi }g_{\mu \nu }=\xi ^{\rho }\partial _{\rho }g_{\mu \nu }+g_{\mu
\rho }\partial _{\nu }\xi ^{\rho }+g_{\nu \rho }\partial _{\mu }\xi ^{\rho }
\label{diffe}
\end{equation}%
denote the transformation of the metric tensor $g_{\mu \nu }$ under
infinitesimal changes of coordinates $\delta x^{\mu }=-\xi ^{\mu }(x)$. The
closure relations read%
\begin{equation}
\lbrack \delta _{\xi },\delta _{\eta }]g_{\mu \nu }=\delta _{\lbrack \xi
,\eta ]}g_{\mu \nu },\qquad \lbrack \xi ,\eta ]^{\rho }\equiv \eta ^{\sigma
}\partial _{\sigma }\xi ^{\rho }-\xi ^{\sigma }\partial _{\sigma }\eta
^{\rho }.  \label{clo1}
\end{equation}

We define the basic cloud field as an independent \textquotedblleft
vector\textquotedblright\ $\zeta ^{\mu }(x)$ that transforms according to
the rule%
\begin{equation}
\delta _{\xi }\zeta ^{\mu }(x)=\xi ^{\mu }(x-\zeta (x)).  \label{dLU}
\end{equation}%
In the next sections we explain how to include $\zeta ^{\mu }$ into the
action. For the moment, we just study its properties. A relation like (\ref%
{dLU}) and similar ones below can be meant as expansions in powers of $\zeta 
$. From now on, we understand that the argument of a function is $x$,
whenever it is not specified.

It is easy to check that the definition (\ref{dLU}) is meaningful, since it
closes:%
\begin{equation}
\lbrack \delta _{\xi },\delta _{\eta }]\zeta ^{\mu }=-\xi ^{\rho }(x-\zeta
)\eta _{,\rho }^{\mu }(x-\zeta )+\eta ^{\rho }(x-\zeta )\xi _{,\rho }^{\mu
}(x-\zeta )=\delta _{\lbrack \xi ,\eta ]}\zeta ^{\mu },  \label{clo2}
\end{equation}%
where $X_{,\mu }\equiv \partial _{\mu }X$. To avoid a certain confusion that
may arise when the argument of a function is $x-\zeta (x)$, we need to pay
attention to the notation. An expression line $\partial _{\rho }X^{\mu
}(x-\zeta )$ is ambiguous, because the total derivative acts on the $x$
dependence inside $\zeta ^{\nu }$, while the partial derivative is not
supposed to. We have 
\begin{equation}
\partial _{\rho }(X^{\mu }(x-\zeta ))=X_{,\rho }^{\mu }(x-\zeta )-\zeta
_{,\rho }^{\sigma }X_{,\sigma }^{\mu }(x-\zeta ).  \label{dex}
\end{equation}

The point-dependent dressed fields are then%
\begin{eqnarray}
\varphi _{\text{d}}(x) &=&\varphi (x-\zeta (x)),\qquad A_{\mu \text{d}%
}(x)=A_{\mu }(x-\zeta )-\zeta _{,\mu }^{\nu }A_{\nu }(x-\zeta ),  \notag \\
g_{\mu \nu \text{d}}(x) &=&g_{\mu \nu }(x-\zeta )-\zeta _{,\mu }^{\rho
}g_{\nu \rho }(x-\zeta )-\zeta _{,\nu }^{\rho }g_{\mu \rho }(x-\zeta )+\zeta
_{,\mu }^{\rho }\zeta _{,\nu }^{\sigma }g_{\rho \sigma }(x-\zeta ),
\label{dressedfields}
\end{eqnarray}%
for scalars, vectors and the metric, respectively. Indeed, using the Taylor
expansion of $\varphi (x-\zeta )$, it is easy to check that $\delta _{\xi
}\varphi =\xi ^{\rho }\varphi _{,\rho }$ implies 
\begin{equation*}
\delta _{\xi }\varphi _{\text{d}}(x)=\xi ^{\rho }(x-\zeta )\varphi _{,\rho
}(x-\zeta )-\delta _{\xi }\zeta ^{\rho }\varphi _{,\rho }(x-\zeta )=0.
\end{equation*}%
Moreover, $\delta _{\xi }A_{\mu }=\xi ^{\rho }A_{\mu ,\rho }+A_{\rho }\xi
_{,\mu }^{\rho }$ implies 
\begin{equation*}
\delta _{\xi }A_{\mu \text{d}}(x)=A_{\rho }(x-\zeta )\xi _{,\mu }^{\rho
}(x-\zeta )-\zeta _{,\mu }^{\nu }A_{\rho }(x-\zeta )\xi _{,\nu }^{\rho
}(x-\zeta )-A_{\rho }(x-\zeta )\partial _{\mu }(\xi ^{\rho }(x-\zeta ))=0,
\end{equation*}%
where we have used (\ref{dex}) in the last step with $X^{\mu }=\xi ^{\mu }$.
Similarly, (\ref{diffe}) implies $\delta _{\xi }g_{\mu \nu \text{d}}(x)=0$.

Generically, if $T_{\mu _{1}\cdots \mu _{n}}(x)$ is a tensor, its gauge
invariant, dressed version is%
\begin{equation}
T_{\mu _{1}\cdots \mu _{n}\hspace{0.01in}\text{d}}(x)=(\delta _{\mu
_{1}}^{\nu _{1}}-\zeta _{,\mu _{1}}^{\nu _{1}})\cdots (\delta _{\mu
_{n}}^{\nu _{n}}-\zeta _{,\mu _{n}}^{\nu _{n}})T_{\nu _{1}\cdots \nu
_{n}}(x-\zeta ).  \label{gendress}
\end{equation}

We can also define dual fields, which allow us to raise and lower the
indices and invert the definitions of the dressed fields given above. The
dual cloud field $\tilde{\zeta}^{\mu }(x)$ is defined as the solution of the
equation%
\begin{equation}
\tilde{\zeta}^{\mu }(x)=-\zeta ^{\mu }(x-\tilde{\zeta}(x)),  \label{dual}
\end{equation}%
which can be worked out recursively by expanding in powers of $\zeta ^{\mu }$%
:%
\begin{equation*}
\tilde{\zeta}^{\mu }(x)=-\zeta ^{\mu }(x+\zeta (x+\zeta (x+\zeta (x+\cdots
)))).
\end{equation*}%
Differentiating (\ref{dual}), we find%
\begin{equation}
\left[ \delta _{\rho }^{\mu }-\zeta _{,\rho }^{\mu }(x-\tilde{\zeta}(x))%
\right] \left[ \delta _{\nu }^{\rho }-\tilde{\zeta}_{,\nu }^{\rho }(x)\right]
=\delta _{\nu }^{\mu }.  \label{jac}
\end{equation}%
Using this identity and (\ref{dLU}) we derive the\ infinitesimal
transformation of $\tilde{\zeta}^{\mu }$, which reads%
\begin{equation}
\delta _{\xi }\tilde{\zeta}^{\mu }(x)=-(\delta _{\nu }^{\mu }-\tilde{\zeta}%
_{,\nu }^{\mu }(x))\xi ^{\nu }(x).  \label{xid}
\end{equation}%
It is straightforward to check its closure.

The inverse relations are%
\begin{eqnarray}
\varphi (x) &=&\varphi _{\text{d}}(x-\tilde{\zeta}(x)),\qquad A_{\mu
}(x)=A_{\mu \text{d}}(x-\tilde{\zeta})-\tilde{\zeta}_{,\mu }^{\nu }A_{\nu 
\text{d}}(x-\tilde{\zeta}),  \notag \\
g_{\mu \nu }(x) &=&g_{\mu \nu \text{d}}(x-\tilde{\zeta})-\tilde{\zeta}_{,\mu
}^{\rho }g_{\nu \rho \text{d}}(x-\tilde{\zeta})-\tilde{\zeta}_{,\nu }^{\rho
}g_{\mu \rho \text{d}}(x-\tilde{\zeta})+\tilde{\zeta}_{,\mu }^{\rho }\tilde{%
\zeta}_{,\nu }^{\sigma }g_{\rho \sigma \text{d}}(x-\tilde{\zeta}),
\label{invertdress} \\
T_{\mu _{1}\cdots \mu _{n}\hspace{0.01in}}(x) &=&(\delta _{\mu _{1}}^{\nu
_{1}}-\tilde{\zeta}_{,\mu _{1}}^{\nu _{1}})\cdots (\delta _{\mu _{n}}^{\nu
_{n}}-\tilde{\zeta}_{,\mu _{n}}^{\nu _{n}})T_{\nu _{1}\cdots \nu _{n}\text{d}%
}(x-\tilde{\zeta}).  \notag
\end{eqnarray}%
Observe that (\ref{dual}) implies%
\begin{equation*}
\left. x^{\mu }-\zeta ^{\mu }(x)\right\vert _{x\rightarrow x-\tilde{\zeta}%
}=x^{\mu }-\tilde{\zeta}^{\mu }(x)-\zeta ^{\mu }(x-\tilde{\zeta}(x))=x^{\mu
}.
\end{equation*}%
Defining $y^{\mu }=x^{\mu }-\tilde{\zeta}^{\mu }(x)$ and relabelling $%
x\leftrightarrow y$, we find the dual identity%
\begin{equation}
\zeta ^{\mu }(x)=-\tilde{\zeta}^{\mu }(x-\zeta (x)).  \label{dualid}
\end{equation}%
Differentiating this relation, we\ also find%
\begin{equation}
\left[ \delta _{\rho }^{\mu }-\tilde{\zeta}_{,\rho }^{\mu }(x-\zeta )\right] %
\left[ \delta _{\nu }^{\rho }-\zeta _{,\nu }^{\rho }(x)\right] =\delta _{\nu
}^{\mu }.  \label{jac2}
\end{equation}

Vectors and tensors with upper indices are dressed as follows: 
\begin{eqnarray}
A_{\text{d}}^{\mu }(x) &=&A^{\mu }(x-\zeta )-\tilde{\zeta}_{,\nu }^{\mu
}(x-\zeta )A^{\nu }(x-\zeta ),  \notag \\
g_{\text{d}\hspace{0.01in}}^{\mu \nu }(x) &=&(\delta _{\rho }^{\mu }-\tilde{%
\zeta}_{,\rho }^{\mu }(x-\zeta ))(\delta _{\sigma }^{\nu }-\tilde{\zeta}%
_{,\sigma }^{\nu }(x-\zeta ))g^{\rho \sigma }(x-\zeta ),
\label{upperdressed} \\
T_{\text{d}\hspace{0.01in}}^{\mu _{1}\cdots \mu _{n}}(x) &=&(\delta _{\nu
_{1}}^{\mu _{1}}-\tilde{\zeta}_{,\nu _{1}}^{\mu _{1}}(x-\zeta ))\cdots
(\delta _{\nu _{n}}^{\mu _{n}}-\tilde{\zeta}_{,\nu _{n}}^{\mu _{n}}(x-\zeta
))T^{\nu _{1}\cdots \nu _{n}}(x-\zeta ).  \notag
\end{eqnarray}%
Indeed, (\ref{jac}) with $x\rightarrow x-\zeta (x)$ ensures that%
\begin{equation*}
A_{\text{d}}^{\mu }(x)A_{\mu \text{d}}(x)=(\delta _{\nu }^{\mu }-\tilde{\zeta%
}_{,\nu }^{\mu }(x-\zeta ))A^{\nu }(x-\zeta )(\delta _{\mu }^{\rho }-\zeta
_{,\mu }^{\rho })A_{\rho }(x-\zeta )=A^{\mu }(x-\zeta )A_{\mu }(x-\zeta ),
\end{equation*}%
as required for a scalar. Similarly, $g_{\text{d}\hspace{0.01in}}^{\mu \nu
}(x)g_{\nu \rho \text{d}\hspace{0.01in}}(x)=\delta _{\rho }^{\mu }$. \
Moreover, the behaviors of upper indices under infinitesimal
transformations, as in $\delta _{\xi }A^{\mu }=\xi ^{\rho }A_{,\rho }^{\mu
}-A^{\rho }\xi _{,\rho }^{\mu }$, imply that the fields (\ref{upperdressed})
are invariant. For example, 
\begin{equation*}
\delta _{\xi }A_{\text{d}}^{\mu }(x)=-(\delta _{\nu }^{\mu }-\tilde{\zeta}%
_{,\nu }^{\mu }(x-\zeta ))\xi _{,\rho }^{\nu }(x-\zeta )A^{\rho }(x-\zeta
)+(\delta _{\nu }^{\mu }-\tilde{\zeta}_{,\nu }^{\mu }(x-\zeta ))\xi _{,\rho
}^{\nu }(x-\zeta )A^{\rho }(x-\zeta )=0.
\end{equation*}%
Here we have used%
\begin{equation*}
\delta _{\xi }\tilde{\zeta}_{,\rho }^{\mu }(x-\zeta )=-(\delta _{\nu }^{\mu
}-\tilde{\zeta}_{,\nu }^{\mu }(x-\zeta ))\xi _{,\rho }^{\nu }(x-\zeta ),
\end{equation*}%
which follows from (\ref{xid}).

It is also crucial to introduce anticommuting partners $H^{\mu }$ of $\zeta
^{\mu }$, defined by the transformation law%
\begin{equation}
\delta _{\xi }H^{\mu }=-H^{\nu }\xi _{,\nu }^{\mu }(x-\zeta ).  \label{dH}
\end{equation}%
The consistency of this transformation follows from its closure:%
\begin{eqnarray}
\lbrack \delta _{\xi },\delta _{\eta }]H^{\mu } &=&H^{\rho }\xi _{,\rho
}^{\nu }(x-\zeta )\eta _{,\nu }^{\mu }(x-\zeta )+H^{\nu }\xi ^{\rho
}(x-\zeta )\eta _{,\nu \rho }^{\mu }(x-\zeta )-(\xi \leftrightarrow \eta ) 
\notag \\
&=&-H^{\rho }(x)(\partial _{\rho }(\eta ^{\nu }\xi _{,\nu }^{\mu }-\xi ^{\nu
}\eta _{,\nu }^{\mu }))(x-\zeta )=\delta _{\lbrack \xi ,\eta ]}H^{\mu }.
\label{ddH}
\end{eqnarray}%
The anticommuting partner $\tilde{H}^{\mu }$ of $\tilde{\zeta}^{\mu }$ is a
field transforming exactly as $H^{\mu }$.

We have achieved what we wanted, that is to say, define point-dependent
observables in general relativity. However, we have done it at the cost of
introducing new fields, the cloud fields (and their anticommuting partners).
The next problem is to include the extra fields into the action, and ensure
that the extension does not change the fundamental theory, and does not
propagate unphysical degrees of freedom. First, we develop a formalism to
ensure that the correlation functions of the undressed fields and the $S$
matrix amplitudes are unmodified, despite the presence of new interactions.
Then, we render the whole new sectors purely virtual.

We also want the construction to be perturbative (expanding the metric
around flat space), diagrammatic and local. We do not require polynomiality,
though, since in quantum gravity we have to renounce it anyway.

\section{Batalin-Vilkovisky formalism and Zinn-Justin master equation}

\setcounter{equation}{0}\label{Bata}

In this section we recall the standard Batalin-Vilkovisky formalism \cite{BV}
for gravity, which is a convenient tool to study the
Ward-Takahashi-Slavnov-Taylor identities \cite{WTST} to all orders in a
compact form.

The classical action $S_{\text{cl}}(g,A,\varphi )$ can be any action of
classical gravity, possibly coupled to matter. For concreteness, we assume
that the matter sector is made of an Abelian vector $A_{\mu }$ and a neutral
scalar field $\varphi $. The specific form of $S_{\text{cl}}$ is not
important for the theoretical setup we are going to develop. However,
particular forms of $S_{\text{cl}}$ will be used in the computations.

We introduce the set of fields $\Phi ^{\alpha }=(g_{\mu \nu },C^{\mu },\bar{C%
}^{\mu },B^{\mu },A_{\mu },\varphi )$, where $C^{\mu }$ are the
Faddeev-Popov ghosts \cite{FP}, $\bar{C}^{\mu }$ are the antighosts and $%
B^{\mu }$ are the Nakanishi-Lautrup Lagrange multipliers \cite{naka}. The
superscript $\alpha $ collects all the indices. We couple sources $K^{\alpha
}=(K_{g}^{\mu \nu },K_{\mu }^{C},K_{\mu }^{\bar{C}},K_{\mu }^{B},$ $%
K_{A}^{\mu }{,}K_{\varphi })$ to the field transformations by means of the
functional%
\begin{eqnarray}
S_{K}(\Phi ,K) &=&-\int (C^{\rho }\partial _{\rho }g_{\mu \nu }+g_{\mu \rho
}\partial _{\nu }C^{\rho }+g_{\nu \rho }\partial _{\mu }C^{\rho })K_{g}^{\mu
\nu }-\int (C^{\rho }\partial _{\rho }A_{\mu }+A_{\rho }\partial _{\mu
}C^{\rho })K_{A}^{\mu }  \notag \\
&&-\int (C^{\rho }\partial _{\rho }\varphi )K_{\varphi }-\int C^{\rho
}(\partial _{\rho }C^{\mu })K_{\mu }^{C}-\int B^{\mu }K_{\mu }^{\bar{C}}.
\label{SK}
\end{eqnarray}%
This way, the transformations of the fields can be written as%
\begin{equation}
\delta _{\xi }\Phi ^{\alpha }=\theta (S_{K},\Phi ^{\alpha })=-\theta \frac{%
\delta _{r}S_{K}}{\delta K^{\alpha }},  \label{gaugetr}
\end{equation}%
where $\xi ^{\mu }=\theta C^{\mu }$, $\theta $ is a constant anticommuting
(Grassmann) variable and%
\begin{equation}
(X,Y)=\int \left( \frac{\delta _{r}X}{\delta \Phi ^{\alpha }}\frac{\delta
_{l}Y}{\delta K^{\alpha }}-\frac{\delta _{r}X}{\delta K^{\alpha }}\frac{%
\delta _{l}Y}{\delta \Phi ^{\alpha }}\right)   \label{brackets}
\end{equation}%
are the Batalin-Vilkovisky antiparentheses \cite{BV}, the subscripts $r$ and 
$l$ denoting the right and left derivatives, respectively.

The closure of the algebra of the transformations is encoded in the identity 
\begin{equation}
(S_{K},S_{K})=0.  \label{closure}
\end{equation}%
The Jacobi identity satisfied by the antiparentheses implies the nilpotence
relation $(S_{K},(S_{K},$ $X))=0$ for every $X$.

The gauge-fixed action reads%
\begin{equation}
S_{\text{gf}}(\Phi )=S_{\text{cl}}(g,A,\varphi )+(S_{K},\Psi (\Phi )),
\label{Sgf}
\end{equation}%
where $\Psi (\Phi )$ is the \textquotedblleft gauge
fermion\textquotedblright , that is to say, a local functional that is
introduced to fix the gauge. For example, in a generic covariant gauge we
may choose%
\begin{equation}
\Psi (\Phi )=\int \sqrt{-g}\bar{C}^{\mu }\left( G_{\mu }(g)-\lambda g_{\mu
\nu }B^{\nu }\right) ,\qquad G_{\mu }(g)=g^{\nu \rho }\partial _{\rho
}g_{\mu \nu }-\frac{\lambda ^{\prime }}{2}g^{\nu \rho }\partial _{\mu
}g_{\nu \rho },  \label{gferm}
\end{equation}%
where $\lambda $, $\lambda ^{\prime }$ are gauge-fixing parameters and $%
G_{\nu }(g)$ is the gauge-fixing function. We have 
\begin{eqnarray*}
(S_{K},\Psi ) &=&\int \sqrt{-g}B^{\mu }\left( G_{\mu }(g)-\lambda g_{\mu \nu
}B^{\nu }\right) +S_{\text{ghost}}\rightarrow \frac{1}{4\lambda }\int \sqrt{%
-g}G_{\mu }g^{\mu \nu }G_{\nu }+S_{\text{ghost}}, \\
S_{\text{ghost}} &=&-\int \bar{C}^{\mu }\left( S_{K},\sqrt{-g}G_{\mu
}-\lambda \sqrt{-g}g_{\mu \nu }B^{\nu }\right) ,
\end{eqnarray*}%
where the arrow denotes the integration over $B^{\mu }$ and $S_{\text{ghost}%
} $ is the ghost action. Other gauge choices will be considered in the paper.

The total action is%
\begin{equation}
S(\Phi ,K)=S_{\text{gf}}(\Phi )+S_{K}(\Phi ,K)  \label{action}
\end{equation}%
and satisfies the Zinn-Justin equation \cite{ZJ} 
\begin{equation}
(S,S)=0,  \label{master}
\end{equation}%
also known as master equation. This identity collects the gauge invariance
of the classical action, the triviality of the gauge-fixing sector, as well
as the closure of the algebra. We have the nilpotence relation $(S,(S,X))=0$
for every $X$.

\section{Cloud sector}

\setcounter{equation}{0}\label{BataCloud}

In this section we build the cloud sector. To trivialize its effects on the
usual correlation functions and the $S$ matrix elements, we mimick the key
aspects of the gauge-fixing procedure. In particular, we need:

1) the cloud field $\zeta ^{\mu }$;

2) its anticommuting partner $H^{\mu }$;

3)\ a new symmetry (which we call cloud symmetry), which shifts $\zeta ^{\mu
}$ by (minus) $H^{\mu }$; and

4) anticommuting $H^{\mu }$-partners $\bar{H}^{\mu }$, as well as Lagrange
multipliers $E^{\mu }$.

\noindent\ The reason why $\zeta ^{\mu }$ and $H^{\mu }$ must have opposite
statistics is precisely that the latter is the shift of the former by the
new symmetry. The reason why $\bar{H}^{\mu }$ and $E^{\mu }$ must be
included is that they allow us to \textquotedblleft fix the
could\textquotedblright , in the same way as we normally fix the gauge.
Indeed, the fields $H^{\mu }$ can be seen as the \textquotedblleft
Faddeev-Popov ghosts\textquotedblright\ of the cloud symmetry, while $\bar{H}%
^{\mu }$ are the \textquotedblleft cloud antighosts\textquotedblright , and $%
E^{\mu }$ are the Lagrange multipliers for the \textquotedblleft
cloud-fixing\textquotedblright . Finally, the reason why $\zeta ^{\mu }$
alone is not sufficient, but $H^{\mu }$, $\bar{H}^{\mu }$ and $E^{\mu }$ are
needed as well, is that the multiplet $\tilde{\Phi}^{\alpha }=(\zeta ^{\mu
},H^{\mu },\bar{H}^{\mu },E^{\mu })$ provides the easiest way to ensure that
the contributions of the extra fields mutually compensate in all the usual,
on-shell correlation functions and the $S$ matrix amplitudes. This way, the
fundamental physics does not change, and the cloud sector makes a difference
only in the new correlations functions, which are those built with the gauge
invariant fields.

Then, we further extend the construction to a whole Batalin-Vilkovisky
formalism for the cloud sector. First, we include a new set of sources $%
\tilde{K}^{\alpha }=(\tilde{K}_{\mu }^{\zeta },\tilde{K}_{\mu }^{H},\tilde{K}%
_{\mu }^{\bar{H}},\tilde{K}_{\mu }^{E})$, coupled to the $\tilde{\Phi}%
^{\alpha }$ transformations. Second, we extend the definition (\ref{brackets}%
) of antiparentheses to the new sector:%
\begin{equation*}
(X,Y)=\int \left( \frac{\delta _{r}X}{\delta \Phi ^{\alpha }}\frac{\delta
_{l}Y}{\delta K^{\alpha }}-\frac{\delta _{r}X}{\delta K^{\alpha }}\frac{%
\delta _{l}Y}{\delta \Phi ^{\alpha }}+\frac{\delta _{r}X}{\delta \tilde{\Phi}%
^{\alpha }}\frac{\delta _{l}Y}{\delta \tilde{K}^{\alpha }}-\frac{\delta _{r}X%
}{\delta \tilde{K}^{\alpha }}\frac{\delta _{l}Y}{\delta \tilde{\Phi}^{\alpha
}}\right) .
\end{equation*}%
Third, we collect the gauge transformations (\ref{dLU}) and (\ref{dH}) of
the new fields $\zeta ^{\mu }$ and $H^{\mu }$, and the cloud
transformations, into the functionals 
\begin{eqnarray*}
S_{K}^{\text{gauge}} &=&S_{K}-\int C^{\mu }(x-\zeta )\tilde{K}_{\mu }^{\zeta
}-\int H^{\nu }C\,_{,\nu }^{\mu }(x-\zeta )\tilde{K}_{\mu }^{H}, \\
S_{K}^{\text{cloud}} &=&\int H^{\mu }\tilde{K}_{\mu }^{\zeta }-\int E^{\mu }%
\tilde{K}_{\mu }^{\bar{H}},\qquad S_{K}^{\text{tot}}=S_{K}^{\text{gauge}%
}+S_{K}^{\text{cloud}}.
\end{eqnarray*}

The cloud transformations, which are encoded into the second functional, are
just the most general shifts of $\zeta ^{\mu }$ and $\bar{H}$. For example,
the total (gauge plus cloud) transformation of $\zeta ^{\mu }$ reads 
\begin{equation*}
\delta _{\xi ,\mathcal{H}}\zeta ^{\mu }=\theta \left( S_{K}^{\text{tot}%
},\zeta ^{\mu }\right) =-\theta H^{\mu }(x)+\theta C^{\mu }(x-\zeta (x))=-%
\mathcal{H}^{\mu }(x)+\xi ^{\mu }(x-\zeta (x)),
\end{equation*}%
where $\mathcal{H}=\theta H$.

It is easy to check the identities%
\begin{equation}
(S_{K}^{\text{gauge}},S_{K}^{\text{gauge}})=(S_{K}^{\text{cloud}},S_{K}^{%
\text{cloud}})=0,  \label{ss1}
\end{equation}%
which express the closures of both types of transformations. The first
identity follows from (\ref{clo1}), (\ref{ddH}) and (\ref{closure}).

We also have:%
\begin{equation}
S_{K}^{\text{gauge}}-S_{K}=-\left( S_{K}^{\text{cloud}},\int C^{\mu
}(x-\zeta )\tilde{K}_{\mu }^{H}\right) ,\qquad S_{K}^{\text{cloud}}=-\left(
S_{K}^{\text{cloud}},\int \zeta ^{\mu }\tilde{K}_{\mu }^{\zeta }+\int \bar{H}%
^{\mu }\tilde{K}_{\mu }^{\bar{H}}\right) .  \label{cexact}
\end{equation}%
These formulas show that the functionals $S_{K}^{\text{gauge}}-S_{K}$ and $%
S_{K}^{\text{cloud}}$ are cohomologically exact under the cloud symmetry,
i.e., they have the form $(S_{K}^{\text{cloud}},$ local functional$)$.
Together with $(S_{K},S_{K}^{\text{cloud}})=0$ (which is trivial), they
imply the further identity 
\begin{equation}
(S_{K}^{\text{gauge}},S_{K}^{\text{cloud}})=0,  \label{Smix}
\end{equation}%
which gives, together with (\ref{ss1}),%
\begin{equation}
(S_{K}^{\text{tot}},S_{K}^{\text{tot}})=0.  \label{ss2}
\end{equation}

\subsection{The cloud and the total action}

We fix the cloud by adding $(S_{K}^{\text{tot}},\tilde{\Psi})$ to the
action, where\ $\tilde{\Psi}(\Phi ,\tilde{\Phi})$ is the \textquotedblleft
cloud fermion\textquotedblright . A typical form of it is\ 
\begin{equation}
\tilde{\Psi}(\Phi ,\tilde{\Phi})=\int \sqrt{-g_{\text{d}}}\bar{H}^{\mu
}\left( V_{\mu }-\tilde{\lambda}g_{\mu \nu \text{d}}E^{\nu }\right) ,
\label{dressed fermion}
\end{equation}%
where $g_{\text{d}}$ is the determinant of $g_{\mu \nu \text{d}}$ and $%
V_{\mu }$ denotes the \textquotedblleft cloud function\textquotedblright ,
i.e., the function that specifies the cloud. We assume that $V_{\mu }$ is
gauge invariant:%
\begin{equation}
(S_{K}^{\text{gauge}},V_{\mu })=0,\qquad (S_{K}^{\text{gauge}},\tilde{\Psi}%
)=0.  \label{spsit}
\end{equation}%
Basically, we can view $V_{\mu }$ as a function of the dressed metric $%
g_{\mu \nu \text{d}}$.

We find%
\begin{eqnarray}
(S_{K}^{\text{tot}},\tilde{\Psi}) &=&(S_{K}^{\text{cloud}},\tilde{\Psi}%
)=\int \sqrt{-g_{\text{d}}}E^{\mu }\left( V_{\mu }-\tilde{\lambda}g_{\mu \nu 
\text{d}}E^{\nu }\right)  \notag \\
&&+\int \bar{H}^{\mu }\frac{\delta }{\delta \zeta ^{\rho }}\left[ \sqrt{-g_{%
\text{d}}}\left( V_{\mu }-\tilde{\lambda}g_{\mu \nu \text{d}}E^{\nu }\right) %
\right] H^{\rho }.  \label{scloud}
\end{eqnarray}%
The last term gives a \textquotedblleft Faddeev-Popov
determinant\textquotedblright\ for the cloud, which is crucial for the
diagrammatic properties that we derive in the next sections.

\medskip

The total action of the extended theory is then 
\begin{equation}
S_{\text{tot}}(\Phi ,K,\tilde{\Phi},\tilde{K})=S_{\text{cl}}+(S_{K}^{\text{%
tot}},\Psi +\tilde{\Psi})+S_{K}^{\text{tot}}  \label{stota}
\end{equation}%
and satisfies its own master equation 
\begin{equation}
(S_{\text{tot}},S_{\text{tot}})=0.  \label{cloudmaster}
\end{equation}%
Note that $S_{\text{tot}}$ is gauge invariant, since equations (\ref{ss1}), (%
\ref{ss2}) and (\ref{spsit}) imply%
\begin{equation}
(S_{K}^{\text{gauge}},S_{\text{tot}})=(S_{K}^{\text{cloud}},S_{\text{tot}%
})=0.  \label{stot}
\end{equation}

We can also write%
\begin{equation*}
S_{\text{tot}}(\Phi ,K,\tilde{\Phi},\tilde{K})=S(\Phi ,K)+(S_{K}^{\text{cloud%
}},\Theta ),\qquad \Theta =\tilde{\Psi}-\int C^{\mu }(x-\zeta )\tilde{K}%
_{\mu }^{H}-\int \zeta ^{\mu }\tilde{K}_{\mu }^{\zeta }-\int \bar{H}^{\mu }%
\tilde{K}_{\mu }^{\bar{H}},
\end{equation*}%
which shows that the difference between the dressed action and the ordinary
action is cohomologically exact with respect to the cloud symmetry.

\subsection{Covariant cloud}

To make explicit calculations, we need to choose the could function $V_{\mu
} $ in (\ref{dressed fermion}). A convenient starting point is the covariant
cloud function 
\begin{equation}
V_{\mu }(g,\zeta )=g_{\text{d}}^{\nu \rho }\partial _{\rho }g_{\mu \nu \text{%
d}}-\frac{\tilde{\lambda}^{\prime }}{2}g_{\text{d}}^{\nu \rho }\partial
_{\mu }g_{\nu \rho \text{d}},  \label{cloud}
\end{equation}%
where $\tilde{\lambda}^{\prime }$ is a further cloud parameter. This choice
mimicks the gauge-fixing function of formula (\ref{gferm}), and makes the
cloud sector as similar as possible to the gauge-fixing one.

Other choices are considered in the paper (see formula (\ref{specialgf}) and
comments right below it). It should be kept in mind that the freedom we have
for $V_{\mu }$ is enormous, given that it describes the influence of the
surrounding, classical environment on the observation of a quantum
phenomenon. When we are interested in a renormalizable theory of quantum
gravity, like the one of \cite{LWgrav}\ (see section \ref{pvc}), it is
convenient to restrict to gauge-fixing functions and cloud functions that
are manifestly renormalizable by power counting. In this paper, we focus on
those.

Note that we are allowed to use second metrics, inside the clouds (as well
as inside the gauge-fixing function). For example, we can use the flat-space
metric $\eta _{\mu \nu }$ to raise the indices of $\partial _{\mu }$.
Sometimes, however, it may be convenient to use a unique metric everywhere,
for a better control on the renormalization properties of the dressed theory.

\section{Cloud independence of the non-cloud sector}

\setcounter{equation}{0}\label{cloudindep}

In this section we prove that the ordinary correlation functions of
elementary and composite fields are unmodified. This also ensures that the
vertices and diagrams of the cloud sector do not affect the renormalization
of the non-cloud sector of the theory. Moreover, we show that the $S$ matrix
amplitudes of the dressed fields are cloud independent and coincide with the
usual $S$ matrix amplitudes of the undressed fields.

The generating functional of the correlation functions is%
\begin{equation}
Z(J,K,\tilde{J},\tilde{K})=\int [\mathrm{d}\Phi \mathrm{d}\tilde{\Phi}]%
\mathrm{\exp }\left( iS_{\text{tot}}(\Phi ,K,\tilde{\Phi},\tilde{K})+i\int
\Phi ^{\alpha }J_{\alpha }+i\int \tilde{\Phi}^{\alpha }\tilde{J}_{\alpha
}\right)  \label{Z}
\end{equation}%
and $W(J,K,\tilde{J},\tilde{K})=-i\ln Z(J,K,\tilde{J},\tilde{K})$ is the
generating functional of the connected ones. The ordinary correlation
functions are the functional derivatives with respect to the sources $%
J^{\alpha }$, calculated at $\tilde{J}=\tilde{K}=0$. They are collected in 
\begin{eqnarray}
Z(J,K,0,0) &=&\int [\mathrm{d}\Phi \mathrm{d}\tilde{\Phi}]\mathrm{\exp }%
\left( iS(\Phi ,K)+i(S_{K}^{\text{tot}},\tilde{\Psi})+i\int \Phi ^{\alpha
}J_{\alpha }\right)  \notag \\
&=&\int [\mathrm{d}\Phi ]\mathrm{\exp }\left( iS(\Phi ,K)+i\int \Phi
^{\alpha }J_{\alpha }\right) \int [\mathrm{d}\tilde{\Phi}]\mathrm{e}%
^{i(S_{K}^{\text{cloud}},\tilde{\Psi})}.  \label{Z0}
\end{eqnarray}%
We want to prove that this expression coincides with the ordinary generating
functional, thanks to the identity%
\begin{equation}
\int [\mathrm{d}\tilde{\Phi}]\mathrm{e}^{i(S_{K}^{\text{cloud}},\tilde{\Psi}%
)}=1.  \label{ide}
\end{equation}

Since $\tilde{\Psi}$ depends on both $\Phi $ and $\tilde{\Phi}$, the
left-hand side of (\ref{ide}) is in principle a functional of $\Phi $. To
show that it is actually a constant, we consider arbitrary infinitesimal
deformations of the fields $\Phi $. Let $\delta \tilde{\Psi}$ denote the
variation of $\tilde{\Psi}$ due to them. The variation of the integral is
then 
\begin{equation}
\delta \int [\mathrm{d}\tilde{\Phi}]\mathrm{e}^{i(S_{K}^{\text{cloud}},%
\tilde{\Psi})}=i\int [\mathrm{d}\tilde{\Phi}](S_{K}^{\text{cloud}},\delta 
\tilde{\Psi})\mathrm{e}^{i(S_{K}^{\text{cloud}},\tilde{\Psi})},
\label{varia}
\end{equation}%
Performing the change of field variables $\tilde{\Phi}^{\alpha }\rightarrow 
\tilde{\Phi}^{\alpha }+\theta (S_{K}^{\text{cloud}},\tilde{\Phi}^{\alpha })$
in the integral%
\begin{equation*}
\int [\mathrm{d}\tilde{\Phi}]\hspace{0.01in}\delta \tilde{\Psi}\hspace{0.01in%
}\mathrm{e}^{i(S_{K}^{\text{cloud}},\tilde{\Psi})},
\end{equation*}%
we obtain%
\begin{equation}
\int [\mathrm{d}\tilde{\Phi}]\hspace{0.01in}\delta \tilde{\Psi}\hspace{0.01in%
}\mathrm{e}^{i(S_{K}^{\text{cloud}},\tilde{\Psi})}=\int [\mathrm{d}\tilde{%
\Phi}]\left[ \delta \tilde{\Psi}+\theta (S_{K}^{\text{cloud}},\delta \tilde{%
\Psi})\right] \mathrm{e}^{i(S_{K}^{\text{cloud}},\tilde{\Psi})}.
\label{cloudin}
\end{equation}%
We have used the fact that $(S_{K}^{\text{cloud}},\tilde{\Psi})$ is
independent of the sources, so $(S_{K}^{\text{cloud}},\tilde{\Psi}%
)\rightarrow (S_{K}^{\text{cloud}},\tilde{\Psi})+\theta (S_{K}^{\text{cloud}%
},(S_{K}^{\text{cloud}},\tilde{\Psi}))=(S_{K}^{\text{cloud}},\tilde{\Psi})$.
The equality (\ref{cloudin}) shows that the right-hand side of (\ref{varia})
vanishes, as we wished to prove.

\subsection{Cloud independence of the \emph{S} matrix amplitudes}

Now we prove that the scattering amplitudes of the dressed fields coincide
with the usual scattering amplitudes (of undressed fields). Specifically,
the clouds have no effect on the mass shell, when the polarizations are
attached to the amputated external legs.

First, we recall a general result about the invariance of the $S$ matrix
amplitudes under a perturbative change of field variables (see, for example, 
\cite{AbsoPhys} for the proof). Consider a generic theory of scalar fields $%
\varphi $, described by some classical action $S(\varphi )$. If $\mathcal{O}%
(\varphi )$ denotes a composite field that is at least quadratic in $\varphi 
$, define new fields $\varphi ^{\prime }\equiv \varphi +\mathcal{O}(\varphi
) $.

Then, the following results hold. The locations $p^{2}=m_{\text{ph}}^{2}$ of
the poles of the two-point functions $\langle \varphi \hspace{0.01in}|%
\hspace{0.01in}\varphi \rangle $ and $\langle \varphi ^{\prime }\hspace{%
0.01in}|\hspace{0.01in}\varphi ^{\prime }\rangle $ coincide perturbatively,
where $m_{\text{ph}}$ denotes the physical mass (possibly equipped with an
imaginary part, if the particle is unstable). We have%
\begin{equation*}
\langle \varphi \hspace{0.01in}|\hspace{0.01in}\varphi \rangle \simeq \frac{%
iZ}{p^{2}-m_{\text{ph}}^{2}+i\epsilon },\qquad \langle \varphi ^{\prime }%
\hspace{0.01in}|\hspace{0.01in}\varphi ^{\prime }\rangle \simeq \frac{%
iZ^{\prime }}{p^{2}-m_{\text{ph}}^{2}+i\epsilon },
\end{equation*}%
for $p^{2}\simeq m_{\text{ph}}^{2}$, for suitable factors $Z$ and $Z^{\prime
}$.

Moreover, the correlation functions that contain more than two $\varphi
^{\prime }$ insertions satisfy%
\begin{equation}
\langle \prod\limits_{a=1}^{j}\frac{p_{a}^{2}-m_{\text{ph}}^{2}}{\sqrt{%
Z^{\prime }}}\varphi ^{\prime }(p_{a})\hspace{0.01in}\rangle _{\text{on-shell%
}}=\langle \hspace{0.01in}\prod\limits_{a=1}^{j}\frac{p_{a}^{2}-m_{\text{ph}%
}^{2}}{\sqrt{Z}}\varphi (p_{a})\rangle _{\text{on-shell}},  \label{onshellco}
\end{equation}%
which proves that the $S$ matrix amplitudes are invariant under the change
of field variables $\varphi \rightarrow \varphi ^{\prime }$.

Applying this theorem to the could extension of the gravity theory, we have,
in momentum space,%
\begin{eqnarray}
&&\langle \prod\limits_{i=1}^{n}k_{i}^{2}\varepsilon _{i\text{d}}^{\mu
_{i}\nu _{i}}(k_{i})g_{\mu _{i}\nu _{i}\text{d}}(k_{i})\prod%
\limits_{a=1}^{j}p_{a}^{2}\varepsilon _{a\text{d}}^{\rho _{a}}(p_{a})A_{\rho
_{a}\text{d}}(p_{a})\prod\limits_{b=1}^{l}\frac{q_{b}^{2}-m_{\text{ph}}^{2}}{%
\sqrt{Z_{\varphi }^{\prime }}}\varphi _{\text{d}}(q_{b})\rangle _{\text{%
on-shell}}  \notag \\
&&\qquad =\langle \prod\limits_{i=1}^{n}k_{i}^{2}\varepsilon _{i}^{\mu
_{i}\nu _{i}}(k_{i})g_{\mu _{i}\nu
_{i}}(k_{i})\prod\limits_{a=1}^{j}p_{a}^{2}\varepsilon _{a}^{\rho
_{a}}(p_{a})A_{\rho _{a}}(p_{a})\prod\limits_{b=1}^{l}\frac{q_{b}^{2}-m_{%
\text{ph}}^{2}}{\sqrt{Z_{\varphi }}}\varphi (q_{b})\rangle _{\text{on-shell}%
}.  \label{onshecorr}
\end{eqnarray}%
The polarizations $\varepsilon ^{\mu \nu }(k)$ and $\varepsilon ^{\mu }(p)$
of the gravitons and the vector fields, respectively, satisfy $k_{\mu
}\varepsilon ^{\mu \nu }(k)=p_{\mu }\varepsilon ^{\mu }(p)=0$, and include
the normalization factors $1/\sqrt{Z}$. The \textquotedblleft
dressed\textquotedblright\ polarizations $\varepsilon _{\text{d}}^{\mu \nu
}(k)$ and $\varepsilon _{\text{d}}^{\mu }(p)$ are the same, apart from
having normalization factors $1/\sqrt{Z^{\prime }}$. With an abuse of
notation, we use the same symbols for the fields and their Fourier
transforms, since the meaning is clear from the context. By the theorem
proved in the first part of this section, the right-hand side of (\ref%
{onshecorr}) is cloud independent and coincides with the usual $S$ matrix
amplitude.

Note that formulas (\ref{dressedfields}) show that the expansion of $g_{\mu
\nu \text{d}}$ in powers of $\zeta ^{\rho }$, combined with the expansion of 
$g_{\mu \nu }$ around the flat-space metric $\eta _{\mu \nu }$, contains a
linear contribution $-\zeta _{\mu ,\nu }-\zeta _{\nu ,\mu }$, besides $%
g_{\mu \nu }$ itself, plus nonlinear terms (which can be regarded as
composite fields $\mathcal{O}(\varphi)$), where $\zeta _{\mu }=\eta \,_{\mu
\nu }\zeta ^{\nu }$. Thus, $g_{\mu \nu \text{d}}$ is not of the form $%
\varphi ^{\prime }=\varphi +\mathcal{O}(\varphi )$. Nevertheless, the linear
terms $-\zeta _{\mu ,\nu }-\zeta _{\nu ,\mu }$ are killed by the
polarization $\varepsilon ^{\mu \nu }(k)$, after\ the Fourier transform.
This means that $\varepsilon _{\text{d}}^{\mu \nu }(k)g_{\mu \nu \text{d}%
}(k) $ is of the required form, apart from an unimportant normalization
factor.

Also note that we are not comparing correlation functions of the same
theory, as in (\ref{onshellco}). We are jumping from one theory (the
extended one, to which the left-hand side of (\ref{onshecorr}) refers) to
another theory (the non extended one, to which the rigth-hand side of (\ref%
{onshecorr}) refers), thanks to the result proved previously in this section.

In the end, the product $k^{2}\varepsilon _{\text{d}}^{\mu \nu }(k)g_{\mu
\nu \text{d}}(k)$ is gauge invariant (and, therefore, gauge independent, for
the arguments given below) and its dressing is trivial: 
\begin{equation*}
\lim_{k^{2}\rightarrow 0}k^{2}\varepsilon _{\text{d}}^{\mu \nu }(k)\langle
g_{\mu \nu \text{d}}(k)\cdots \rangle =\lim_{k^{2}\rightarrow
0}k^{2}\varepsilon ^{\mu \nu }(k)\langle g_{\mu \nu }(k)\cdots \rangle .
\end{equation*}%
The same result holds for the insertions of vectors and scalars (and
fermions, if present).

The identity (\ref{onshecorr}) proves that the ordinary theory of scattering
can be understood as a theory of scattering of dressed fields. We can even
forget about the undressed fields altogether, and always work with the
dressed fields. So doing, gauge invariance and gauge independence become
manifest. In particular, the $S$-matrix amplitudes are automatically ensured
to be gauge independent.

Clearly, the proof of this section relies heavily on the notion of
asymptotic state, which is crucial to build the $S$ matrix elements. For
this reason, it does not generalize to short-distance scattering processes,
where the incoming and outgoing states are not allowed to become free. Those
processes can be studied from the correlation functions of gauge-invariant
fields (which are not constrained to be on the mass shell), once they are
equipped with the notion of pure virtuality.

\section{Dressed correlation functions}

\label{cloudcorre}\setcounter{equation}{0}

In this section we study the correlation functions that contain insertions
of dressed fields. It is possible to study them systematically by coupling
new sources to them and extending the generating functionals again. We
replace the action $S_{\text{tot}}$ inside (\ref{Z}) by%
\begin{equation}
S_{\text{tot}}^{\text{ext}}=S_{\text{tot}}+\int \left( J_{\text{d}}^{\mu \nu
}g_{\mu \nu \text{d}}+J_{\text{d}}^{\mu }A_{\mu \text{d}}+J_{\text{d}%
}\varphi _{\text{d}}\right) ,  \label{extra}
\end{equation}%
and denote the extended functionals by $Z_{\text{tot}}^{\text{ext}}(J,K,%
\tilde{J},\tilde{K},J_{\text{d}})=\exp (iW_{\text{tot}}^{\text{ext}}(J,K,%
\tilde{J},\tilde{K},J_{\text{d}}))$. The insertions of dressed fields can be
studied by taking the functional derivatives with respect to the new sources 
$J_{\text{d}}^{\mu \nu },J_{\text{d}}^{\mu },J_{\text{d}}$. The extended
action is gauge invariant, since (\ref{stot}) implies%
\begin{equation}
(S_{K}^{\text{gauge}},S_{\text{tot}}^{\text{ext}})=0.  \label{stotext}
\end{equation}%
Clearly, $S_{\text{tot}}^{\text{ext}}$ is not cloud invariant.

It is straightforward to prove that the correlation functions of the dressed
fields, collected in the functional $Z_{\text{tot}}^{\text{ext}}(J_{\text{d}%
})=\exp (iW_{\text{tot}}^{\text{ext}}(J_{\text{d}}))\equiv Z_{\text{tot}}^{%
\text{ext}}(0,0,0,0,J_{\text{d}})=\exp (iW_{\text{tot}}^{\text{ext}}(0,0,0,$ 
$0,J_{\text{d}}))$, are gauge independent. The argument is identical to the
one of subsection 7.1 of \cite{AbsoPhys}, so we do not repeat it here. Gauge
independence will be verified explicitly in the computations.

\section{Gauge/cloud duality}

\setcounter{equation}{0}\label{duality}

In this section we show that the gauge-trivial sector and the cloud sector
are dual to each other. We call this property \textit{gauge/cloud duality}.
Sometimes, it can be used to simplify the computations.

We begin by noting that the transformation law (\ref{dLU}) and the
definitions (\ref{dressedfields}) imply that the could transformation of the
dressed metric is just an infinitesimal diffeomorphism (\ref{diffe}) with
parameters $H_{\text{d}}^{\mu }$ such that%
\begin{equation}
H^{\mu }=(\delta _{\nu }^{\mu }-\zeta _{,\nu }^{\mu })H_{\text{d}}^{\nu }.
\label{Hd}
\end{equation}%
Precisely, 
\begin{equation}
(S_{K}^{\text{tot}},g_{\mu \nu \text{d}})=(S_{K}^{\text{cloud}},g_{\mu \nu 
\text{d}})=H_{\text{d}}^{\rho }\partial _{\rho }g_{\mu \nu \text{d}}+g_{\mu
\rho \text{d}}\partial _{\nu }H_{\text{d}}^{\rho }+g_{\nu \rho \text{d}%
}\partial _{\mu }H_{\text{d}}^{\rho }=\delta _{H_{\text{d}}}^{\text{diff}%
}g_{\mu \nu \text{d}}.  \label{SAd}
\end{equation}%
Similarly, for scalars and vectors we have%
\begin{equation}
(S_{K}^{\text{cloud}},\varphi _{\text{d}})=H_{\text{d}}^{\rho }\partial
_{\rho }\varphi _{\text{d}},\qquad (S_{K}^{\text{cloud}},A_{\mu \text{d}%
})=H_{\text{d}}^{\rho }\partial _{\rho }A_{\mu \text{d}}+A_{\rho \text{d}%
}\partial _{\mu }H_{\text{d}}^{\rho }.  \label{SAd2}
\end{equation}%
It is easy to check that $H_{\text{d}}^{\mu }$ is gauge invariant $(S_{K}^{%
\text{gauge}},H_{\text{d}}^{\mu })=0$. Moreover, the cloud transformation of 
$H_{\text{d}}^{\mu }$ mimics the gauge transformation of the ghosts $C^{\mu
} $:%
\begin{equation}
(S_{K}^{\text{cloud}},H_{\text{d}}^{\mu })=H_{\text{d}}^{\rho }\partial
_{\rho }H_{\text{d}}^{\mu }.  \label{SHd}
\end{equation}

We define the dressed cloud field $\zeta _{\text{d}}^{\mu }$ as the dual
field of formula (\ref{dual}):%
\begin{equation}
\zeta _{\text{d}}^{\mu }=\tilde{\zeta}^{\mu }.  \label{dual cloud}
\end{equation}%
Using (\ref{Hd}), it is easy to derive the cloud transformation of $\zeta _{%
\text{d}}^{\mu }$, which reads%
\begin{equation}
(S_{K}^{\text{cloud}},\zeta _{\text{d}}^{\mu })=H_{\text{d}}^{\mu }(x-\zeta
_{\text{d}}(x)).  \label{Szd}
\end{equation}%
Note that $\zeta _{\text{d}}^{\mu }$ is not gauge invariant. Using (\ref{xid}%
), its gauge transformation can be used to define the dressed Faddeev-Popov
ghosts%
\begin{equation}
C_{\text{d}}^{\mu }\equiv (\delta _{\nu }^{\mu }-\zeta _{\text{d},\nu }^{\mu
})C^{\nu }=-(S_{K}^{\text{gauge}},\zeta _{\text{d}}^{\mu }),  \label{Cd}
\end{equation}%
which, instead, are gauge invariant by construction. Their cloud
transformations read%
\begin{equation}
(S_{K}^{\text{cloud}},C_{\text{d}}^{\mu })=(S_{K}^{\text{gauge}},(S_{K}^{%
\text{cloud}},\zeta _{\text{d}}^{\mu }))=(S_{K}^{\text{gauge}},H_{\text{d}%
}^{\mu }(x-\zeta _{\text{d}}(x)))=C_{\text{d}}^{\rho }H_{,\rho \text{d}%
}^{\mu }(x-\zeta _{\text{d}}),  \label{SKCd}
\end{equation}%
having used (\ref{Smix}) and $(S_{K}^{\text{gauge}},H_{\text{d}}^{\mu })=0$.

Now, collecting (\ref{dressedfields}), (\ref{Hd}), (\ref{dual cloud}) and (%
\ref{Cd}), we define the change of field variables%
\begin{equation}
\Phi ,\tilde{\Phi}\rightarrow \Phi _{\text{d}},\tilde{\Phi}_{\text{d}}
\label{redefa}
\end{equation}%
from undressed fields to dressed fields, leaving all the other fields
unchanged: $\bar{C}_{\text{d}}^{\mu }=\bar{C}^{\mu }$, $B_{\text{d}}^{\mu
}=B^{\mu }$, $\bar{H}_{\text{d}}^{\mu }=\bar{H}^{\mu }$ and $E_{\text{d}}=E$%
. The transformations are perturbatively local, which means that when we use
them as changes of field variables in the functional integral, the Jacobian
determinant is equal to one, using the dimensional regularization.

To ensure that the antiparentheses are preserved, so that all the properties
derived till now continue to hold, we embed (\ref{redefa}) into a canonical
transformation%
\begin{equation}
\Phi ,\tilde{\Phi},K,\tilde{K}\rightarrow \Phi _{\text{d}},\tilde{\Phi}_{%
\text{d}},K_{\text{d}},\tilde{K}_{\text{d}},  \label{canonical}
\end{equation}%
of the Batalin-Vilkovisky\ type. Its generating functional is 
\begin{equation*}
F(\Phi ,\tilde{\Phi},K_{\text{d}},\tilde{K}_{\text{d}})=\int \Phi _{\text{d}%
}(\Phi ,\tilde{\Phi})K_{\text{d}}+\int \tilde{\Phi}_{\text{d}}(\Phi ,\tilde{%
\Phi})\tilde{K}_{\text{d}}.
\end{equation*}

At the practical level, the whole operation amounts to work out the
transformations of the dressed fields, which we have already done, and
couple them to the dressed sources. Collecting the gauge transformations (%
\ref{Cd}) and the cloud transformations (\ref{SAd}), (\ref{SAd2}), (\ref{SHd}%
), (\ref{Szd}) and (\ref{SKCd}), we find 
\begin{eqnarray*}
S_{K}^{\text{gauge}} &=&\int C_{\text{d}}^{\mu }\tilde{K}_{\mu \text{d}%
}^{\zeta }-\int B_{\text{d}}^{\mu }K_{\mu \text{d}}^{\bar{C}}, \\
S_{K}^{\text{cloud}} &=&-\int (H_{\text{d}}^{\rho }\partial _{\rho }g_{\mu
\nu \text{d}}+g_{\mu \rho \text{d}}\partial _{\nu }H_{\text{d}}^{\rho
}+g_{\nu \rho \text{d}}\partial _{\mu }H_{\text{d}}^{\rho })K_{g\text{d}%
}^{\mu \nu }-\int H_{\text{d}}^{\rho }(\partial _{\rho }H_{\text{d}}^{\mu
})K_{\mu \text{d}}^{H} \\
&&-{\int H_{\text{d}}^{\rho }(\partial _{\rho }\varphi _{\text{d}%
})K_{\varphi \text{d}}-\int (H_{\text{d}}^{\rho }\partial _{\rho }A_{\text{d}%
\mu }+A_{\rho \text{d}}\partial _{\mu }H_{\text{d}}^{\rho })K_{\mu \text{d}%
}^{A}} \\
&&-\int E_{\text{d}}^{\mu }\tilde{K}_{\mu \text{d}}^{\bar{H}}-\int H_{\text{d%
}}^{\mu }(x-\zeta _{\text{d}})\tilde{K}_{\mu \text{d}}^{\zeta }-\int C_{%
\text{d}}^{\nu }H_{,\nu \text{d}}^{\mu }(x-\zeta _{\text{d}})\tilde{K}_{\mu 
\text{d}}^{C}.
\end{eqnarray*}

We see that the canonical transformation (\ref{canonical}) switches the
gauge transformations and the cloud transformations. Similarly, it exchanges
the roles of the gauge-fixing function $G_{\mu }$ and the cloud function $%
V_{\mu }$: $G_{\mu }(g(\zeta ,g_{\text{d}}))\leftrightarrow V_{\mu }(g_{%
\text{d}})$. We may also say that it exchanges the quantization prescription
of the gauge-trivial sector with the one of the cloud sector (see below).

The correlation functions of the dressed fields coincide with the ones of
the undressed fields in a specific gauge. For example, choosing the
covariant gauge (\ref{gferm}) and the covariant cloud (\ref{dressed fermion}%
), (\ref{cloud}), we have%
\begin{eqnarray}
&&\langle g_{\mu _{1}\nu _{1}\text{d}}(x_{1})\cdots \hspace{0.01in}g_{\mu
_{n}\nu _{n}\text{d}}(x_{n})\hspace{0.01in}\varphi _{\text{d}}(y_{1})\cdots
\varphi _{\text{d}}(y_{j})\hspace{0.01in}A_{\rho _{1}\text{d}}(z_{1})\cdots
A_{\rho _{k}\text{d}}(z_{k})\rangle  \notag \\
&&\qquad =\langle g_{\mu _{1}\nu _{1}}(x_{1})\cdots \hspace{0.01in}g_{\mu
_{n}\nu _{n}}(x_{n})\hspace{0.01in}\varphi (y_{1})\cdots \varphi (y_{j})%
\hspace{0.01in}A_{\rho _{1}}(z_{1})\cdots A_{\rho _{k}}(z_{k})\rangle
_{\lambda \rightarrow \tilde{\lambda},\lambda ^{\prime }\rightarrow \tilde{%
\lambda}^{\prime }}.  \label{lr}
\end{eqnarray}%
Combined with the cloud independence of the right-hand side, proved in
section \ref{cloudindep}, this property ensures that the dressed correlation
function can be calculated by replacing $\lambda $ with $\tilde{\lambda}$
and $\lambda ^{\prime }$ with $\tilde{\lambda}^{\prime }$ in a usual
correlation function. The left-hand side of (\ref{lr}) normally includes a
huge number of diagrams. However, (\ref{lr}) implies\ that most
contributions cancel out in the end.

\section{Multiclouds}

\setcounter{equation}{0}\label{multiclouds}

In this section we show how to equip each insertion with its own,
independent dressing. To do so, we extend the formalism of the previous
sections by adding several copies of the could sector.

We introduce many cloud fields $\zeta ^{\mu i}$, where $i$ labels the
copies. Then we add copies of their anticommuting partners $H^{\mu i}$ (the
cloud ghosts), the antighosts $\bar{H}^{\mu i}$ and the Lagrange multipliers 
$E^{\mu i}$. We collect them in $\tilde{\Phi}^{\alpha i}=(\zeta ^{\mu
i},H^{\mu i},\bar{H}^{\mu i},E^{\mu i})$. We also couple sources $\tilde{K}%
^{\alpha i}$ to their transformations. Next, we extend the definition (\ref%
{brackets}) of antiparentheses to include all the copies:%
\begin{equation}
(X,Y)=\int \left[ \frac{\delta _{r}X}{\delta \Phi ^{\alpha }}\frac{\delta
_{l}Y}{\delta K^{\alpha }}-\frac{\delta _{r}X}{\delta K^{\alpha }}\frac{%
\delta _{l}Y}{\delta \Phi ^{\alpha }}+\sum_{i}\left( \frac{\delta _{r}X}{%
\delta \tilde{\Phi}^{\alpha i}}\frac{\delta _{l}Y}{\delta \tilde{K}^{\alpha
i}}-\frac{\delta _{r}X}{\delta \tilde{K}^{\alpha i}}\frac{\delta _{l}Y}{%
\delta \tilde{\Phi}^{\alpha i}}\right) \right] .  \label{BVi}
\end{equation}%
Finally, we extend the gauge transformations and introduce cloud
transformations for each copy: 
\begin{eqnarray}
S_{K}^{\text{gauge}} &=&S_{K}-\sum_{i}\int C^{\mu }(x-\zeta ^{i})\tilde{K}%
_{\mu }^{\zeta i}-\sum_{i}\int H^{\nu i}C\,_{,\nu }^{\mu }(x-\zeta ^{i})%
\tilde{K}_{\mu }^{Hi},  \notag \\
S_{K}^{\text{cloud\hspace{0.01in}}i} &=&\int (H^{\mu i}\tilde{K}_{\mu
}^{\zeta i}-E^{\mu i}\tilde{K}_{\mu }^{\bar{H}i}),\qquad S_{K}^{\text{cloud}%
}=\sum_{i}S_{K}^{\text{cloud\hspace{0.01in}}i},\qquad S_{K}^{\text{tot}%
}=S_{K}^{\text{gauge}}+S_{K}^{\text{cloud}}.\qquad  \label{Scloudi}
\end{eqnarray}%
It is easy to check that the identities (\ref{ss1}) and (\ref{ss2}) continue
to hold.

The simplest cloud fermion is just the sum of the cloud fermions of each
copy:\ 
\begin{equation}
\tilde{\Psi}(\Phi ,\tilde{\Phi})=\sum_{i}\int \sqrt{-g_{\text{d}}^{i}}\bar{H}%
^{\mu i}\left( V_{\mu }^{i}+\frac{\tilde{\lambda}_{i}}{2}g_{\mu \nu \text{d}%
}^{i}E^{\nu i}\right) ,  \label{psiti}
\end{equation}%
where $g_{\mu \nu \text{d}}^{i}$ is the dressed metric tensor built with the 
$i$th cloud field $\zeta ^{\mu i}$, and $V_{\mu }^{i}$ is the $i$th cloud
function, assumed to be gauge invariant, $(S_{K}^{\text{gauge}},V_{\mu
}^{i})=0$. For simplicity, we also assume that each $V_{\mu }^{i}$ depends
only on the $i$th cloud field $\zeta ^{\mu i}$ (besides $g_{\mu \nu }$),
i.e., different cloud sectors are not mixed. We can just take each $V_{\mu
}^{i}$ to be a function of $g_{\mu \nu \text{d}}^{i}$.

The total action of the extended theory is still (\ref{stota}),\ and
satisfies the master equations (\ref{cloudmaster}) and (\ref{stot}).
Moreover,%
\begin{equation}
(S_{K}^{\text{cloud\hspace{0.01in}}i},S_{\text{tot}})=0  \label{SKtoti}
\end{equation}%
for every $i$.

It is always possible to build gauge invariant functions with two cloud
fields. For example, the functions%
\begin{equation}
\zeta _{1i}^{\mu }(x)\equiv \zeta ^{\mu 1}(x)+\tilde{\zeta}^{\mu i}(x-\zeta
^{1}(x))  \label{ginvcl}
\end{equation}%
are gauge invariant, since (\ref{dLU}) and (\ref{xid}) imply%
\begin{equation*}
\delta _{\xi }\zeta _{1i}^{\mu }(x)=\xi ^{\mu }(x-\zeta ^{1})-\tilde{\zeta}%
_{,\nu }^{\mu i}(x-\zeta ^{1})\xi ^{\mu }(x-\zeta ^{1})-(\delta _{\nu }^{\mu
}-\tilde{\zeta}_{,\nu }^{\mu i}(x-\zeta ^{1}))\xi ^{\nu }(x-\zeta ^{1})=0.
\end{equation*}%
We do not have control on such functions, when they are turned on. For this
reason, it may be important to prove, when possible, that the operations we
make preserve the unmixing stated above.

The insertions of dressed fields can be studied by means of the extended
action%
\begin{equation}
S_{\text{tot}}^{\text{ext}}=S_{\text{tot}}+\sum_{i}\int \left( J_{\text{d}%
}^{\mu \nu i}g_{\mu \nu \text{d}}^{i}+J_{\text{d}}^{\mu i}A_{\mu \text{d}%
}^{i}+J_{\text{d}}^{i}\varphi _{\text{d}}^{i}\right) .  \label{sexti}
\end{equation}

The correlation functions that do not contain insertions belonging to some
cloud sector are independent of that cloud sector. Indeed, the proof of (\ref%
{ide}) can be repeated for every cloud sector separately.

The gauge/cloud duality is less powerful in the presence of many clouds. It
can be used to eliminate one cloud, or a combination of clouds, but not all
of them. For example, a correlation function 
\begin{equation}
\langle g_{\mu _{1}\nu _{1}\text{d}}^{(1)}(x_{1})\cdots \hspace{0.01in}%
g_{\mu _{n}\nu _{n}\text{d}}^{(n)}(x_{n})\hspace{0.01in}A_{\rho _{1}\text{d}%
}^{(n+1)}(y_{1})\cdots A_{\rho _{j}\text{d}}^{(n+j)}(y_{j})\hspace{0.01in}%
\varphi _{\text{d}}^{(n+j+1)}(z_{1})\cdots \varphi _{\text{d}%
}^{(n+j+k)}(z_{k})\rangle ,  \label{dre}
\end{equation}%
with different clouds for every field, can be simplified to%
\begin{equation}
\langle g_{\mu _{1}\nu _{1}}(x_{1})g_{\mu _{2}\nu _{2}\text{d}}^{(2)\hspace{%
0.01in}\prime }(x_{1})\cdots \hspace{0.01in}g_{\mu _{n}\nu _{n}\text{d}}^{(n)%
\hspace{0.01in}\prime }(x_{n})\hspace{0.01in}A_{\rho _{1}\text{d}}^{(n+1)%
\hspace{0.01in}\prime }(y_{1})\cdots A_{\rho _{j}\text{d}}^{(n+j)\hspace{%
0.01in}\prime }(y_{j})\hspace{0.01in}\varphi _{\text{d}}^{(n+j+1)\hspace{%
0.01in}\prime }(z_{1})\cdots \varphi _{\text{d}}^{(n+j+k)\hspace{0.01in}%
\prime }(z_{k})\rangle ,  \notag
\end{equation}%
by means of a field redefinition that exchanges the first dressed field with
its undressed version. The primes mean that the clouds of the other fields
must be redefined as a consequence.

These operations preserve the unmixing, after further redefinitions of the
cloud fields. For example, the transformation (\ref{canonical}) leads to%
\begin{equation*}
\varphi _{\text{d}}^{(i)}(x)=\varphi _{\text{d}}^{(1)}(x-\zeta ^{i\hspace{%
0.01in}\prime }(x)),\qquad i>n+j,
\end{equation*}%
where $\varphi _{\text{d}}^{(1)}$ is the scalar field dressed with the first
cloud (which does not even appear in (\ref{dre}), but this does not matter
for what we are saying) and $\zeta ^{i\hspace{0.01in}\prime }(z_{i})$ is the
solution of%
\begin{equation*}
\zeta ^{\mu i\hspace{0.01in}\prime }(x)=\zeta ^{\mu i}(x)-\zeta ^{\mu
1}(x-\zeta ^{i\hspace{0.01in}\prime }(x)).
\end{equation*}%
To restore the unmixing, it is sufficient to define the new $i$th cloud
field as $\zeta ^{\mu i\hspace{0.01in}\prime }$, $i>1$, after relabelling $%
\varphi _{\text{d}}^{(1)}$ as $\varphi $. The same can be done for the other
insertions of (\ref{dre}).

Repeating the arguments of section \ref{cloudindep}, it is possible to
extend the results of that section to the multicloud case, i.e., prove that
the usual correlation functions are cloud independent, and that the $S$
matrix amplitudes coincide with the usual ones, formula (\ref{onshecorr}),
even when each insertion is dressed with its own, independent cloud.

In the explicit calculations of this paper we work with a unique cloud, for
simplicity.

\section{One-loop two-point functions}

\setcounter{equation}{0}\label{dressed2ptf}

In this section we calculate the one-loop two-point functions of the basic
dressed fields in the covariant gauge, with a covariant cloud. We show that
the absorptive parts are in general unphysical. In the next section we turn
to purely virtual clouds, and show that the absorptive parts then become
physical.

We define the expansion around flat space by writing $g_{\mu \nu }=\eta
_{\mu \nu }+2\kappa h_{\mu \nu }$ and $g_{\mu \nu \text{d}}=\eta _{\mu \nu
}+2\kappa h_{\mu \nu \text{d}}$, where $\kappa =$ $\sqrt{8\pi G}$ and $G$ is
Newton's constant. It is convenient to make the replacements%
\begin{eqnarray*}
C^{\mu } &\rightarrow &\kappa C^{\mu },\qquad B^{\mu }\rightarrow \kappa
B^{\mu },\qquad \bar{C}^{\mu }\rightarrow \kappa \bar{C}^{\mu },\qquad \Psi
\rightarrow \kappa ^{-2}\Psi , \\
\zeta ^{\mu } &\rightarrow &\kappa \zeta ^{\mu },\qquad H^{\mu }\rightarrow
\kappa H^{\mu },\qquad E^{\mu }\rightarrow \kappa E^{\mu },\qquad \bar{H}%
^{\mu }\rightarrow \kappa \bar{H}^{\mu },\qquad \tilde{\Psi}\rightarrow
\kappa ^{-2}\tilde{\Psi},
\end{eqnarray*}%
so that the loop expansion coincides with the expansion in powers of $\kappa 
$.

The two-point functions can be calculated by expanding the dressed fields\ (%
\ref{dressedfields}) to the first order in $\kappa $, where we find%
\begin{eqnarray}
\varphi _{\text{d}} &=&\varphi -\kappa \zeta ^{\mu }\varphi _{,\mu },\qquad
A_{\mu \text{d}}=A_{\mu }-\kappa \zeta ^{\rho }A_{\mu ,\rho }-\kappa \zeta
_{,\mu }^{\rho }A_{\rho },  \notag \\
h_{\mu \nu \text{d}} &=&h_{\mu \nu }-\frac{1}{2}(\zeta _{\mu ,\nu }+\zeta
_{\nu ,\mu })-\kappa \zeta ^{\rho }h_{\mu \nu ,\rho }-\kappa \zeta _{,\mu
}^{\rho }h_{\nu \rho }-\kappa \zeta _{,\nu }^{\rho }h_{\mu \rho }+\frac{%
\kappa }{2}\zeta _{,\mu }^{\rho }\eta _{\rho \sigma }\zeta _{,\nu }^{\sigma
},  \notag
\end{eqnarray}%
where $\zeta _{\mu }=\eta _{\mu \nu }\zeta ^{\nu }$. The higher-order
corrections can be neglected in our calculations, since they give only
tadpoles.

We start from Einstein gravity minimally coupled to a massless scalar field $%
\varphi $ and a vector field $A_{\mu }$. The action is%
\begin{equation*}
-\frac{1}{16\pi G}\int \mathrm{d}^{4}x\sqrt{-g}R+\frac{1}{2}\int \mathrm{d}%
^{4}x\sqrt{-g}g^{\mu \nu }(\partial _{\mu }\varphi )(\partial _{\nu }\varphi
)-\frac{1}{4}\int \mathrm{d}^{4}x\sqrt{-g}F_{\mu \nu }F_{\rho \sigma }g^{\mu
\rho }g^{\nu \sigma }.
\end{equation*}

The two-point function of the dressed scalar field reads%
\begin{equation*}
\langle \varphi _{\text{d}}\hspace{0.01in}|\hspace{0.01in}\varphi _{\text{d}%
}\rangle =\langle \varphi \hspace{0.01in}|\hspace{0.01in}\varphi \rangle
-\kappa \langle \zeta ^{\mu }\varphi _{,\mu }\hspace{0.01in}|\hspace{0.01in}%
\varphi \rangle -\kappa \langle \varphi \hspace{0.01in}|\hspace{0.01in}\zeta
^{\mu }\varphi _{,\mu }\rangle +\kappa ^{2}\langle \zeta ^{\mu }\varphi
_{,\mu }\hspace{0.01in}|\hspace{0.01in}\zeta ^{\nu }\varphi _{,\nu }\rangle +%
\mathcal{O}(\kappa ^{3}),
\end{equation*}%
to the quadratic order in $\kappa $. A vertical bar separates the
(elementary or composite) field of momentum $p$ (to the left) from the one
of momentum $-p$ (to the right). The diagrams are shown in fig. \ref%
{dressedfermion}. 
\begin{figure}[t]
\begin{center}
\includegraphics[width=12truecm]{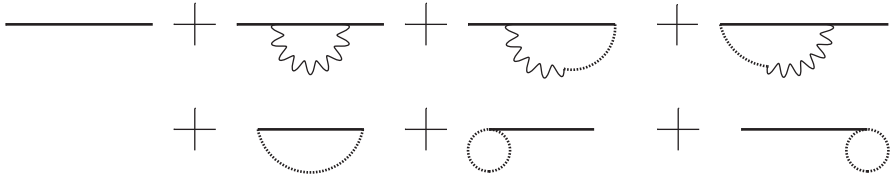}
\end{center}
\caption{Two-point function of the dressed scalar field to order $\protect%
\kappa ^{2}$}
\label{dressedfermion}
\end{figure}

The covariant gauge defined by (\ref{gferm}) gives the ordinary scalar
self-energy%
\begin{equation*}
\langle \varphi \hspace{0.01in}|\hspace{0.01in}\varphi \rangle =\frac{i}{%
p^{2}+i\epsilon }+\frac{3i\kappa ^{2}(\text{$\lambda \lambda ^{\prime }$}%
-3\lambda -\text{$\lambda ^{\prime }$}+5)(\text{$\lambda ^{\prime }$}-1)}{%
16\pi ^{2}\varepsilon (\text{$\lambda ^{\prime }$}-2)^{2}}(-p^{2}-i\epsilon
)^{-\varepsilon /2}+\mathcal{O}(\kappa ^{3}).
\end{equation*}%
Using the covariant cloud (\ref{cloud}), the remaining diagrams of fig. \ref%
{dressedfermion} give the total%
\begin{equation}
\langle \varphi _{\text{d}}\hspace{0.01in}|\hspace{0.01in}\varphi _{\text{d}%
}\rangle =\frac{i}{p^{2}+i\epsilon }+\frac{3i\kappa ^{2}(\text{$\tilde{%
\lambda}\tilde{\lambda}^{\prime }$}-3\text{$\tilde{\lambda}$}-\text{$\tilde{%
\lambda}^{\prime }$}+5)(\text{$\tilde{\lambda}^{\prime }$}-1)}{16\pi
^{2}\varepsilon (\text{$\tilde{\lambda}^{\prime }$}-2)^{2}}(-p^{2}-i\epsilon
)^{-\varepsilon /2}+\mathcal{O}(\kappa ^{3}).  \label{psidpsid}
\end{equation}%
The dependence on the gauge-fixing parameters $\lambda $ and $\lambda
^{\prime }$ has disappeared, as expected. The result depends on the choice
of the cloud, through the parameters $\tilde{\lambda}$ and $\tilde{\lambda}%
^{\prime }$, and satisfies formula (\ref{lr}), due to the gauge/cloud
duality.

The off-shell absorptive part of the two-point function is defined by
amputating the external legs and taking the real part, multiplied by minus 2
(see \cite{AbsoPhys} for details). We find%
\begin{eqnarray*}
\text{Abso}[\langle \varphi _{\text{d}}\hspace{0.01in}|\hspace{0.01in}%
\varphi _{\text{d}}\rangle ] &=&-2\text{Re}[(ip^{2})\langle \varphi _{\text{d%
}}\hspace{0.01in}|\hspace{0.01in}\varphi _{\text{d}}\rangle (ip^{2})] \\
&=&-\frac{3\kappa ^{2}(\text{$\tilde{\lambda}\tilde{\lambda}^{\prime }$}-3%
\text{$\tilde{\lambda}$}-\text{$\tilde{\lambda}^{\prime }$}+5)(\text{$\tilde{%
\lambda}^{\prime }$}-1)}{16\pi (\text{$\tilde{\lambda}^{\prime }$}-2)^{2}}%
(p^{2})^{2}\theta (p^{2})+\mathcal{O}(\kappa ^{3}).
\end{eqnarray*}%
The sign of the lowest-order contribution is positive or negative, depending
on the cloud parameters $\tilde{\lambda}$ and $\tilde{\lambda}^{\prime }$,
so Abso$[\langle \varphi _{\text{d}}\hspace{0.01in}|\hspace{0.01in}\varphi _{%
\text{d}}\rangle ]$ is not physical.

In the case of the vector field, the undressed two-point function reads%
\begin{equation*}
\langle A_{\mu }\hspace{0.01in}|\hspace{0.01in}A_{\nu }\rangle =\langle
A_{\mu }\hspace{0.01in}|\hspace{0.01in}A_{\nu }\rangle _{0}+\frac{i\kappa
^{2}\left( 3\lambda -(1-2\lambda ^{\prime })^{2}\right) }{24\pi
^{2}\varepsilon (\lambda ^{\prime }-2)^{2}}\left( \eta _{\mu \nu }-\frac{%
p_{\mu }p_{\nu }}{p^{2}}\right) (-p^{2}-i\epsilon )^{-\varepsilon /2},
\end{equation*}%
where $\langle A_{\mu }\hspace{0.01in}|\hspace{0.01in}A_{\nu }\rangle _{0}$
is the free propagator. After the gravitational dressing, we find%
\begin{equation*}
\langle A_{\mu \text{d}}\hspace{0.01in}|\hspace{0.01in}A_{\nu \text{d}%
}\rangle =\langle A_{\mu }\hspace{0.01in}|\hspace{0.01in}A_{\nu }\rangle
_{0}+\frac{i\kappa ^{2}\left( 3\tilde{\lambda}-(1-2\tilde{\lambda}^{\prime
})^{2}\right) }{24\pi ^{2}\varepsilon (\tilde{\lambda}^{\prime }-2)^{2}}%
\left[ \eta _{\mu \nu }+f(\lambda ,\lambda ^{\prime },\tilde{\lambda},\tilde{%
\lambda}^{\prime },\lambda _{A})\frac{p_{\mu }p_{\nu }}{p^{2}}\right]
(-p^{2}-i\epsilon )^{-\varepsilon /2},
\end{equation*}%
where $f$ is a function that we do not report here, while $\lambda _{A}$ is
the gauge-fixing parameter of the $A_{\mu }$ propagator. To get rid of $%
\lambda _{A}$, we must include a gauge dressing for $A_{\mu }$ (besides the
gravitational dressings we have already included). This operation is
straightforward, since it is sufficient to consider the two-point function $%
\langle F_{\mu \nu \text{d}}\hspace{0.01in}|\hspace{0.01in}F_{\rho \sigma 
\text{d}}\rangle $ of the field strength. We obtain $\langle F_{\mu \nu 
\text{d}}\hspace{0.01in}|\hspace{0.01in}F_{\rho \sigma \text{d}}\rangle $ $%
=\langle F_{\mu \nu }\hspace{0.01in}|\hspace{0.01in}F_{\rho \sigma }\rangle
_{\lambda \rightarrow \tilde{\lambda},\lambda ^{\prime }\rightarrow \tilde{%
\lambda}^{\prime }}$, in agreement with (\ref{lr}). The absorptive part
reads 
\begin{eqnarray*}
&&\text{Abso}[\langle F_{\mu \nu \text{d}}\hspace{0.01in}|\hspace{0.01in}%
F_{\rho \sigma \text{d}}\rangle ]=-2\text{Re}[(ip^{2})\langle F_{\mu \nu 
\text{d}}\hspace{0.01in}|\hspace{0.01in}F_{\rho \sigma \text{d}}\rangle
(ip^{2})] \\
&&\qquad =-\frac{\kappa ^{2}\left( 3\tilde{\lambda}-(1-2\tilde{\lambda}%
^{\prime })^{2}\right) }{24\pi (\tilde{\lambda}^{\prime }-2)^{2}}%
(p^{2})^{2}\theta (p^{2})(\eta _{\mu \rho }p_{\nu }p_{\sigma }-\eta _{\mu
\sigma }p_{\nu }p_{\rho }-\eta _{\nu \rho }p_{\mu }p_{\sigma }+\eta _{\nu
\sigma }p_{\mu }p_{\rho }).
\end{eqnarray*}%
Again, it is not physical.

\section{Purely virtual clouds and physical absorptive parts}

\setcounter{equation}{0}\label{pvc}

To find physical absorptive parts, we must turn to purely virtual clouds.
This is achieved as follows. The free propagators mix the graviton field $%
h_{\mu \nu }$ and the cloud field $\zeta ^{\mu }$. Inside the propagators,
we can distinguish three types of poles in $p^{2}$: the physical poles, the
gauge-trivial poles and the cloud poles. The cloud poles are those
introduced by the cloud, and appear in $\langle h_{\mu \nu }|\zeta ^{\rho
}\rangle _{0}$ and $\langle \zeta ^{\rho }|\zeta ^{\sigma }\rangle _{0}$.
The gauge-trivial poles are those involving the unphysical components of $%
h_{\mu \nu }$, which are $h_{00}$, $h_{0i}$, the longitudinal components $%
p^{j}h_{ij}(p)$ and the trace $h_{ii}$ (in some reference frame), where $i,j$
are space indices. The physical poles are the remaining ones.

The three classes of poles can be clearly distinguished in the special gauge
of ref. \cite{unitarityc}, which can be extended to the could sector
straightforwardly. The gauge fermion (\ref{gferm}) is replaced by%
\begin{eqnarray}
\Psi (\Phi ) &=&\int \bar{C}^{0}\left( G_{0}(g)-\frac{\lambda +3}{4}%
B_{0}\right) +\int \bar{C}^{i}\left( G_{i}(g)-\lambda \frac{\lambda +3}{4}%
B_{i}\right) ,  \notag \\
G_{0}(g) &=&\lambda \partial _{0}h_{00}+\partial _{0}h_{ii}-2\partial
_{i}h_{0i},  \label{specialgf} \\
G_{i}(g) &=&2\lambda \partial _{0}h_{0i}-\lambda \partial _{i}h_{00}-\frac{%
\lambda +3}{2}\partial _{j}h_{ij}+\frac{\lambda +1}{2}\partial _{i}h_{jj}. 
\notag
\end{eqnarray}%
The cloud fermion (\ref{dressed fermion}) is replaced by an analogous
formula, with $\lambda \rightarrow \tilde{\lambda}$, $\bar{C}^{\mu
}\rightarrow \bar{H}^{\mu }$, $B^{\mu }\rightarrow E^{\mu }$, $G_{\mu
}\rightarrow V_{\mu }$, $h_{\mu \nu }\rightarrow h_{\mu \nu \text{d}}=h_{\mu
\nu }-(\partial _{\mu }\zeta _{\nu }+\partial _{\nu }\zeta _{\mu })/2+%
\mathcal{O}(\kappa )$.

We do not report the free propagators explicitly, because they are quite
lengthy and not strictly necessary for our calculations (see \cite{AbsoPhys}
for their expressions in gauge theories). We just report that they contain
only single poles (which is what makes the special gauge \textquotedblleft
special\textquotedblright ), and that the gauge-trivial poles are located at 
$\lambda E^{2}-\mathbf{p}^{2}=0$ and $4\lambda E^{2}-\mathbf{p}%
^{2}(3+\lambda )=0$, the cloud poles are located at $\tilde{\lambda}E^{2}-%
\mathbf{p}^{2}=0$ and $4\tilde{\lambda}E^{2}-\mathbf{p}^{2}(3+\tilde{\lambda}%
)=0$, and the physical poles are obviously located at $E^{2}-\mathbf{p}%
^{2}=0 $, where $p^{\mu }=(E,\mathbf{p})$ is the propagator momentum. What
is important is that a unique gauge-fixing parameter, $\lambda $, and a
unique cloud parameter, $\tilde{\lambda}$, are sufficient to distinghuish
the three classes of poles in a manifest way.

The cloud poles must be quantized as purely virtual \cite{diagrammarMio}.
This means that, after performing the threshold decomposition of a diagram
as explained in ref. \cite{diagrammarMio}, the (multi) thresholds receiving
contributions from those poles must be removed. The gauge-trivial poles can
be quantized in the way we want (because the correlation functions of the
dressed fields are gauge independent). The physical poles must be quantized
by means of the Feynman $i\epsilon $ prescription.

It is convenient to quantize the gauge-trivial poles as purely virtual as
well, like the cloud poles. Since the purely virtual poles do not contribute
to the absorptive parts of the two-point functions at one loop, we can just
ignore all of them.

At the end, each calculation amounts to just one diagram, the usual
self-energy diagram (second drawing of fig. \ref{dressedfermion}), with a
caveat: we must replace the internal graviton and vector propagators with
their physical parts, which are%
\begin{eqnarray}
\langle A^{\mu a}\hspace{0.01in}|\hspace{0.01in}A^{\nu b}\rangle _{0\text{%
phys}} &=&i\frac{\delta ^{ab}\delta _{i}^{\mu }\delta _{j}^{\nu }\Pi ^{ij}}{%
p^{2}+i\epsilon },  \notag \\
\langle h_{\mu \nu }\hspace{0.01in}|\hspace{0.01in}h_{\rho \sigma }\rangle
_{0\text{phys}} &=&\frac{i}{2}\frac{\delta _{i}^{\mu }\delta _{j}^{\nu
}\delta _{k}^{\rho }\delta _{l}^{\sigma }}{p^{2}+i\epsilon }(\Pi ^{ik}\Pi
^{jl}+\Pi ^{il}\Pi ^{jk}-\Pi ^{ij}\Pi ^{kl}),  \label{physprop}
\end{eqnarray}%
where $\Pi ^{ij}=\delta ^{ij}-(p^{i}p^{j}/\mathbf{p}^{2})$.

In the scalar case, the absorptive part of $\langle \varphi _{\text{d}}%
\hspace{0.01in}|\hspace{0.01in}\varphi _{\text{d}}\rangle $ turns out to be
zero. We can partially understand this result by noting that the optical
theorem relates it to the cross section of a process (graviton emission by a
scalar field), which cannot occur on shell. Nevertheless, the correlation
function we are studying is not on shell. Yet, the result is still zero, due
to the graviton polarizations, which are implicit in (\ref{physprop}).

In the case of the vectors, we find, at rest,%
\begin{equation}
\text{Abso}[\langle A_{i\text{d}}\hspace{0.01in}|\hspace{0.01in}A_{j\text{d}%
}\rangle ]=-2\text{Re}[(ip^{2})\langle A_{i\text{d}}\hspace{0.01in}|\hspace{%
0.01in}A_{j\text{d}}\rangle (ip^{2})]=\frac{\kappa ^{2}\delta _{ij}}{12\pi }%
(p^{2})^{2}\theta (p^{2})+\mathcal{O}(\kappa ^{3}),  \label{absovec}
\end{equation}%
which is positive, as expected.

Finally, the absorptive part of the dressed graviton two-point function (in
pure gravity, at rest) is%
\begin{eqnarray}
\text{Abso}[\langle h_{ij\text{d}}\hspace{0.01in}|\hspace{0.01in}h_{kl\text{d%
}}\rangle ] &=&-2\text{Re}[(ip^{2})\langle h_{ij\text{d}}\hspace{0.01in}|%
\hspace{0.01in}h_{kl\text{d}}\rangle (ip^{2})]  \notag \\
&=&\frac{\kappa ^{2}(3\delta _{ik}\delta _{jl}+3\delta _{il}\delta
_{jk}-2\delta _{ij}\delta _{kl})}{80\pi }(p^{2})^{2}\theta (p^{2})+\mathcal{O%
}(\kappa ^{3}),  \label{absogravo}
\end{eqnarray}%
which is again positive definite.

Now we switch to the theory of quantum gravity with purely virtual particles 
\cite{LWgrav}. It is convenient to formulate it in the variables of ref. 
\cite{AbsoGrav}, to gain an explicit distinction among the graviton, the
inflaton $\phi $ and the massive purely virtual spin-2 particle $\chi _{\mu
\nu }$. We can actually ignore $\chi _{\mu \nu }$, because it does not
contribute to the absorptive parts that we want to compute. Neglecting the
cosmological constant, the relevant terms of the action are%
\begin{eqnarray*}
&&-\frac{1}{16\pi G}\int \mathrm{d}^{4}x\sqrt{-g}R+\frac{1}{2}\int \sqrt{-g}%
\left[ g^{\mu \nu }(\partial _{\mu }\phi )(\partial _{\nu }\phi )-\frac{%
3m_{\phi }^{2}}{2\kappa ^{2}}\left( 1-\mathrm{e}^{\kappa \phi \sqrt{2/3}%
}\right) ^{2}\right] \\
&&+\frac{1}{2}\int \mathrm{d}^{4}x\sqrt{-g}g^{\mu \nu }\mathrm{e}^{\kappa
\phi \sqrt{2/3}}(\partial _{\mu }\varphi )(\partial _{\nu }\varphi )-\frac{1%
}{4}\int \mathrm{d}^{4}x\sqrt{-g}F_{\mu \nu }F_{\rho \sigma }g^{\mu \rho
}g^{\nu \sigma }.
\end{eqnarray*}

The absorptive part (\ref{absovec})\ of the vector two-point function does
not change. The one of the scalar two-point function is no longer zero,
because it receives a contribution from the inflaton. In the high-energy
limit (where we can neglect the mass $m_{\phi }$), we find%
\begin{equation*}
\text{Abso}[\langle \varphi _{\text{d}}\hspace{0.01in}|\hspace{0.01in}%
\varphi _{\text{d}}\rangle ]=\frac{\kappa ^{2}}{48\pi }(p^{2})^{2}\theta
(p^{2})+\mathcal{O}(\kappa ^{3}).
\end{equation*}%
Switching off the matter sector, the absorptive part (\ref{absogravo}) of
the dressed graviton two-point function also receives a correction from the
inflaton $\phi $, and the final result is (\ref{absogravo}) multiplied by
19/18.

\section{Renormalization}

\setcounter{equation}{0}\label{renormalization}

In this section we study the renormalization of the extended theory. We
assume that the starting theory of quantum gravity is renormalizable by
power counting, like the theory based on purely virtual particles of ref. 
\cite{LWgrav}. To have better power-counting behaviors, it may be convenient
to use a higher-derivative gauge-fixing, as in \cite{UVQG}, and
higher-derivative clouds as well. The cloud fields $\zeta ^{\mu i}$ have
dimension minus one in units of mass, so nonpolynomial functions of them are
turned on by renormalization.

The theory of \cite{LWgrav} is unitary. If the clouds are purely virtual,
the complete dressed theory is unitary as well. However, the arguments of
this section do not rely on unitarity, so the results we obtain also apply
to nonunitary clouds, and even nonunitary theories, such as the Stelle
theory \cite{stelle}, where the Feynman prescription is used for the
quantization of the every field (and so $\chi _{\mu \nu }$ is a ghost).

When the arguments work for gravity exactly as they do for gauge theories,
we skip the details of the proofs. The reader should refer to \cite{AbsoPhys}
for the missing derivations.

First, the master equation (\ref{cloudmaster}) satisfied by $S_{\text{tot}}$
implies an analogous master equation%
\begin{equation}
(\Gamma _{\text{tot}},\Gamma _{\text{tot}})=0  \label{stotot}
\end{equation}%
for the generating functional $\Gamma _{\text{tot}}=W_{\text{tot}}(J,\tilde{J%
},K,\tilde{K})-\int \Phi ^{\alpha }J_{\alpha }-\sum_{i}\int \tilde{\Phi}%
^{\alpha i}\tilde{J}_{\alpha }^{i}$ of the one-particle irreducible (1PI)
Green functions, where $\Phi ^{\alpha }=\delta _{r}W_{\text{tot}}/\delta
J^{\alpha }$, $\tilde{\Phi}^{\alpha i}=\delta _{r}W_{\text{tot}}/\delta 
\tilde{J}^{\alpha i}$. Second, the $i$th cloud invariance (\ref{stot}) of
the total action $S_{\text{tot}}$, which is the identity $(S_{K}^{\text{cloud%
\hspace{0.01in}}i},S_{\text{tot}})=0$, implies the $i$th cloud invariance%
\begin{equation}
(S_{K}^{\text{cloud\hspace{0.01in}}i},\Gamma _{\text{tot}})=0
\label{scloudtot}
\end{equation}%
of the $\Gamma $ functional.

Proceeding inductively, we can show that the total renormalized action $S_{R%
\hspace{0.01in}\text{tot}}$ satisfies the renormalized master equations 
\begin{equation}
(S_{R\hspace{0.01in}\text{tot}},S_{R\hspace{0.01in}\text{tot}})=0,\qquad
(S_{K}^{\text{cloud\hspace{0.01in}}i},S_{R\hspace{0.01in}\text{tot}})=0.
\label{SRtotmast}
\end{equation}

Since the cloud symmetry is the most general shift of the cloud fields, the
second equation ensures that the total renormalized action is the sum of a
cloud-independent renormalized action $S_{R}$ and some cloud-exact rest.
Separating $S_{K}^{\text{cloud}}$ itself, which is nonrenormalized, we can
write 
\begin{equation*}
S_{R\hspace{0.01in}\text{tot}}=S_{R}+(S_{K}^{\text{cloud}},\Upsilon
_{R})+S_{K}^{\text{cloud}}
\end{equation*}%
for some local functional $\Upsilon _{R}$. It is possible to extend the
arguments of section \ref{cloudindep} to $S_{R\hspace{0.01in}\text{tot}}$
and show that $S_{R}$ is cloud independent and coincides with the usual
renormalized action, while the scattering amplitudes are gauge independent.

Moreover, the dependence on the gauge-fixing parameters and the dependences
on the cloud parameters go through renormalization as canonical
transformations. In particular, the beta functions of the physical
parameters are gauge independent and cloud independent.

Some simplification comes from the introduction of \textquotedblleft cloud
numbers\textquotedblright , besides the usual ghost number. The usual ghost
number is defined to be equal to 1 for $C^{\mu }$, minus 1 for $\bar{C}^{\mu
}$, $K_{\mu }^{B}$, $K_{g}^{\mu \nu }$, ${K_{A}^{\mu }}$, ${K_{\varphi }}$, $%
\tilde{K}_{\mu }^{\zeta i}$ and $\tilde{K}_{\mu }^{Hi}$, minus 2 for $K_{\mu
}^{C}$, and 0 for every other field and source. The $i$th cloud number is
defined to be equal to one for $H^{\mu i}$, minus one for $\bar{H}^{\mu i}$, 
$\tilde{K}_{\mu }^{Hi}$ and $\tilde{K}_{\mu }^{Ei}$, and zero in all the
other cases.

Every term of the action $S_{\text{tot}}$ is neutral with respect to the
ghost and cloud numbers just defined, with the exception of the source terms 
$\int H^{\mu i}\tilde{K}_{\mu }^{\zeta i}$. Since, however, such terms
cannot be used in nontrivial 1PI diagrams, all the counterterms are neutral.
This ensures that each cloud number is separately conserved in 1PI diagrams
beyond the tree level.

Power counting is not very helpful in the cloud sectors, since the cloud
fields $\zeta ^{\mu i}$ have negative dimensions. A simplification can be
achieved by combining the background-field method with the
Batalin-Vilkovisky formalism, as shown in ref. \cite{NocohoKSZc}. So doing,
the gauge and cloud transformations are not renormalized. Yet, the cloud
sectors are nonrenormalizable, strictly speaking, since infinitely many
counterterms are allowed by power counting and the symmetry constraints. For
example, we can always build gauge invariant candidate counterterms that
depend nontrivially on the fields of each cloud sector and are exact under
every cloud symmetry. Examples are 
\begin{equation*}
(S_{K}^{\text{cloud\hspace{0.01in}}1},(S_{K}^{\text{cloud\hspace{0.01in}}%
2},\ldots (S_{K}^{\text{cloud\hspace{0.01in}}N},\tilde{\Upsilon}))),
\end{equation*}%
where $N$ denotes the number of clouds, and $\tilde{\Upsilon}$ is a gauge
invariant local functional, built with gauge invariant combinations of cloud
fields, such as (\ref{ginvcl}). We cannot exclude that different cloud
sectors mix under renormalization. Nevertheless, we can prove that the
counterterms that do not contain fields and sources of some cloud sector are
the same as if that sector were absent.

The renormalization in every non-background-field approach can be reached by
means of a (renormalized) canonical transformation. Details on this can be
found in ref. \cite{NocohoKSZc}.

Equipped with the renormalized action and the renormalized gauge
transformations, we can build dressed fields that are gauge-invariant with
respect to the latter. The correlation functions of the renormalized dressed
fields are gauge independent.

Finally, the arguments that lead to the identity (\ref{onshecorr}) continue
to hold after renormalization. We have an identity analogous to (\ref%
{onshecorr}), where the dressed and undressed fields are replaced by their
renormalized versions. In particular, the $S$ matrix amplitudes of the
renormalized dressed fields are cloud independent and coincide with the
usual $S$ matrix amplitudes of the renormalized undressed fields. Since the
former are gauge independent by construction, the latter are gauge
independent as well.

\section{Conclusions}

\label{conclusions}\setcounter{equation}{0}

We have extended quantum field theory to include purely virtual cloud
sectors, to study point-dependent physical observables in general relativity
and quantum gravity, with particular emphasis on the gauge invariant
versions of the metric and the matter fields. The cloud diagrammatics and
its Feynman rules are derived from a local action, which is built by means
of cloud fields $\zeta ^{\mu i}$ and their anticommuting partners $H^{\mu i}$%
. It incorporates the choices of clouds, the cloud Faddeev-Popov
determinants and the cloud symmetries. The usual gauge-fixing must be cloud
invariant, while the cloud-fixings must be gauge invariant.

The formalism allows us to define physical, off-mass-shell correlation
functions of point-dependent observables, and calculate them within the
realm of perturbative quantum field theory. Every field insertion can be
equipped with its own, independent cloud. We may eventually replace the
elementary fields with the dressed ones everywhere, to work in a manifestly
gauge independent environment.

The extension does not change the fundamental physics, in the sense that the
ordinary correlation functions and the $S$ matrix amplitudes are unmodified.
It allows us to compute new correlation functions, those containing
insertions of dressed fields. If the clouds are quantized as purely virtual,
the extended theory is unitary. In particular, the correlation functions of
the dressed fields obey the off-shell, diagrammatic version of the optical
theorem. No unwanted degrees of freedom propagate.

A Batalin-Vilkovisky formalism and its Zinn-Justin master equations allow us
to study renormalizability and the WTST identities to all orders in the
perturbative expansion. A gauge/cloud duality shows that the usual
gauge-fixing is nothing but a particular cloud, provided it is rendered
purely virtual. A purely virtual gauge-fixing is a natural upgrade of the
so-called physical gauges \cite{physga}.

We have illustrated the key properties of our approach by computing the
one-loop two-point functions of dressed scalars, vectors and gravitons, and
comparing purely virtual clouds to non-purely virtual clouds, in Einstein
gravity as well as in quantum gravity with purely virtual particles. If
purely virtual clouds are used, the absorptive parts are positive, cloud
independent and gauge independent. This suggests that they are properties of
the fundamental theory. The absorptive parts are not positive, in general,
if non-purely virtual clouds are used.

Pure virtuality can be a natural environment to extract physical information
from off-shell correlation functions. Among the other things, it allows us
to break global invariances without breaking the local ones, avoiding
undesirable consequences on unitarity. We can also define short-distance
scattering processes, where the results depend on the coulds. It emerges
that in such processes the observer necessarily disturbs\ the observed
phenomenon. The net effect is that the amplitudes are physical, but depend
on the cloud parameters. Yet, after sacrificing a few measurements for the
calibration of our instrumentation, we are able to make testable, and
possibly falsifiable, predictions.

We did not introduce true matter to define the metric as a physical
observable. Instead, we used purely virtual dressings. In this sense, our
approach provides the identitication of a complete set of observables in
quantum gravity. Yet, it raises new issues. It would be interesting to
clarify the relation between the Komar-Bergmann classical approach \cite%
{Komar} and the one formulated here, as well as investigate the nonlocal
nature of the algebra of commutators (check \cite{Giddings} for this aspect
in the Donnelly-Giddings approach). It is important to recall that pure
virtuality at the operatorial level still has to be understood (the
formulation we have today being mainly diagrammatic \cite{diagrammarMio}),
so we may not be ready to use the results of this paper for a canonical
analysis of the observables and a Hamiltonian quantization.

The formulation developed here is perturbative, around flat space. To
overcome these limitations, we need to face old and new challenges. An
obstacle is that the prescription for purely virtual particles is understood
only diagrammatically (hence, perturbatively) at present. Even standard
procedures, like the resummation of self-energies into effective
propagators, hide unexpected features, as discussed in ref. \cite{muon}.
Another challenge is to generalize or adapt the known nonperturbative
methods. The numeric (lattice) approaches are not immediately helpful, since
they are mostly suited for Euclidean theories, where the crucial aspects of
purely virtual particles disappear. Results about the perturbative expansion
around a non flat background have been obtained in the context of primordial
cosmology \cite{ABP}. Still, a systematics of purely virtual particles in
curved spacetime is awaiting to be developed.

\vskip.3 truecm \noindent {\large \textbf{Acknowledgments}}

\vskip .5 truecm

This work was supported in part by the European Regional Development Fund
through the CoE program grant TK133 and the Estonian Research Council grant
PRG803. We thank the CERN theory group for hospitality during the final
stage of the project.

\end{document}